\documentclass[aps, prl, twocolumn, nobalancelastpage, amsmath, amssymb, superscriptaddress, nofootinbib]{revtex4-1}

\usepackage{graphicx}
\usepackage{bm}
\usepackage{natbib}
\usepackage{hyperref}
\usepackage{amsmath}
\usepackage{amssymb}
\usepackage{maybemath}
\usepackage{xcolor}
\usepackage{ulem}
\usepackage{mathrsfs}
\usepackage{tikz}

\begin{document}

\title{Improved access to the fine-structure constant with the simplest atomic systems}

\author{H. Cakir}
\email[]{halil.cakir@mpi-hd.mpg.de}
\thanks{This article comprises parts of the Ph.D. thesis work of H.~C. and I.~A.~V. to be submitted to Heidelberg University, Germany.}
\affiliation{Max~Planck~Institute for Nuclear Physics, Saupfercheckweg~1, D~69117 Heidelberg, Germany}

\author{N.~S. Oreshkina}
\affiliation{Max~Planck~Institute for Nuclear Physics, Saupfercheckweg~1, D~69117 Heidelberg, Germany}

\author{I.~A.~Valuev}
\affiliation{Max~Planck~Institute for Nuclear Physics, Saupfercheckweg~1, D~69117 Heidelberg, Germany}

\author{V.~Debierre}
\affiliation{Max~Planck~Institute for Nuclear Physics, Saupfercheckweg~1, D~69117 Heidelberg, Germany}

\author{V.~A. Yerokhin}
\affiliation{Max~Planck~Institute for Nuclear Physics, Saupfercheckweg~1, D~69117 Heidelberg, Germany}
\affiliation{Center for Advanced Studies, Peter the Great St.~Petersburg Polytechnic University, 195251 St.~Petersburg, Russia}

\author{C.~H. Keitel}
\affiliation{Max~Planck~Institute for Nuclear Physics, Saupfercheckweg~1, D~69117 Heidelberg, Germany}

\author{Z. Harman}
\email[]{harman@mpi-hd.mpg.de}
\affiliation{Max~Planck~Institute for Nuclear Physics, Saupfercheckweg~1, D~69117 Heidelberg, Germany}

\begin{abstract}
  A means to extract the fine-structure constant~$\alpha$ from precision
  spectroscopic data on one-electron ions is presented.  We show that in an
  appropriately weighted difference of the bound-electron $g$~factor and the
  ground-state energy, nuclear structural effects can be effectively
  suppressed. This method is anticipated to deliver an independent value of $\alpha$ via existing or near-future
  combined Penning trap and x-ray spectroscopic technology, and enables decreasing the uncertainty
  of $\alpha$ by orders of magnitude.
\end{abstract}

\maketitle

The fine structure constant $\alpha$ is a dimensionless quantity characterizing the strength of the electromagnetic interaction.
Besides photon recoil experiments~\cite{Bouchendira2011,Parker2018}, measurements of the free-electron $g$~factor deliver its most
accurate values~\cite{Hanneke2008,Mohr2012}, with a current standard uncertainty of
$\delta\alpha = 1.1 \cdot 10^{-12}$~\cite{Tiesinga2020}. An independent value of $\alpha$ can be obtained from the $g$~factor of an
electron bound in a single-electron ion of atomic number $Z$: Isolating the leading relativistic
(Dirac) contribution~\cite{Breit1928} $g_{\mathrm{D}} = \frac{2}{3} (2\sqrt{1 - (Z\alpha)^{2}}+1) = 2 - \frac{2}{3}(Z\alpha)^{2} + \dots$,
and calculating QED and nuclear corrections, $\alpha$ can be extracted in principle from the experimental $g$~factor.
$g_{\rm D}$ is most
sensitive to $\alpha$ in highly charged ions (HCI). These ions, however, feature large nuclear structural
effects due to charge radii, nuclear charge distributions and polarizabilities, which are not known with sufficient accuracy, and set
a limitation on the theory of the bound-electron $g$~factor and thus on the accuracy of $\alpha$ determination.

In Ref.~\cite{ShabaevPRL2006}, a specific difference of the $g$~factors of heavy H-like and B-like ions of a heavy element was
put forward to suppress nuclear effects. The combination of light H- and Li-like ions was suggested in Ref.~\cite{YerokhinPRL2016}.
These methods require a significant development of many-electron theory. While important progress has been achieved
(see e.g.~\cite{Glazov2019,Arapoglou2019,Koehler2016,Wagner2013,Yerokhin2017,Cakir2020}),
a further substantial decrease of theoretical uncertainties is needed.

In this Letter we put forward a weighted difference of the $g$~factor and the bound-electron energy $E$ of a
H-like ion (in natural units, $\hbar=c=1$),
\begin{equation}
  \label{eq:defreduced}
  \widetilde{g} \equiv g-x\frac{E}{m_{\rm e}}  \,,
\end{equation}
for which, with an appropriately chosen weight $x$, a strong suppression of nuclear effects can be
achieved. In the above formula, $m_{\rm e}$ is the electron mass. As will be discussed below, the weight is known analytically as
\begin{equation}
  \label{eq:x}
  x=\frac{4}{3} \left(2\sqrt{1 - (Z\alpha)^{2}} + 1\right)\,,
\end{equation}
following from basic properties the Dirac equation~\cite{Karshenboim2005}.
Modern measurements of the $g$~factor of HCI have reached a relative accuracy of $3\times 10^{-11}$~\cite{Sturm2014,Wagner2013,Sturm2011,Sturm2011b,Arapoglou2019}.
It is important to note that $E=m_{\rm e}-E_{\rm B}$ is the total ground-state ($1s$) energy of the electron, i.e. the
rest energy minus the binding energy $E_{\rm B}$, thus $E$ is known to higher relative precision than $E_{\rm B}$ alone.
The ground-state binding energy can be determined e.g. with x-ray spectroscopy
\cite{Decaux1997,Beiersdorfer1989,Tavernier1985,Marmar1986,Briand1984,Deslattes1984,Kubicek2014,Bruhns2007,Rudolph2013,Epp2007,Bernitt2012,GumberidzePRL2005},
using the theoretical value of the excited-state energy.
As an example, with an error bar of 14~meV for the Ly$_{\alpha}$ transition energy of Ar${}^{17+}$~\cite{Kubicek2014},
$E$ can be extracted with a fractional uncertainty of 2.7$\times$10${}^{-8}$.
Very recently, it was demonstrated that electronic binding energies of HCI can also be measured by mass spectrometry~\cite{Rischka2020}.

Accompanied by corresponding foreseeable improvements in the QED description of one-electron systems,
such experiments would become sensitive to the uncertainty of the fine-structure constant $\alpha$.
An important advantage of our scheme based on the reduced $g$ factor~(\ref{eq:defreduced}), compared to those relying on few-electron
ions~\cite{ShabaevPRL2006,YerokhinPRL2016}, is that the theory of one-electron ions is
more advanced than that of many-body systems. Further progress is anticipated to be achieved faster,
rendering decreasing the uncertainty of $\alpha$ by orders of magnitude significantly more likely.
In what follows, we show how the individual nuclear structural terms, such as the leading and QED finite size effects, as well as the
the nuclear polarization contribution, are suppressed in the difference~$\widetilde{g}$.

\begin{figure*}[t!]
  \centering
  \includegraphics[width=0.8\textwidth]{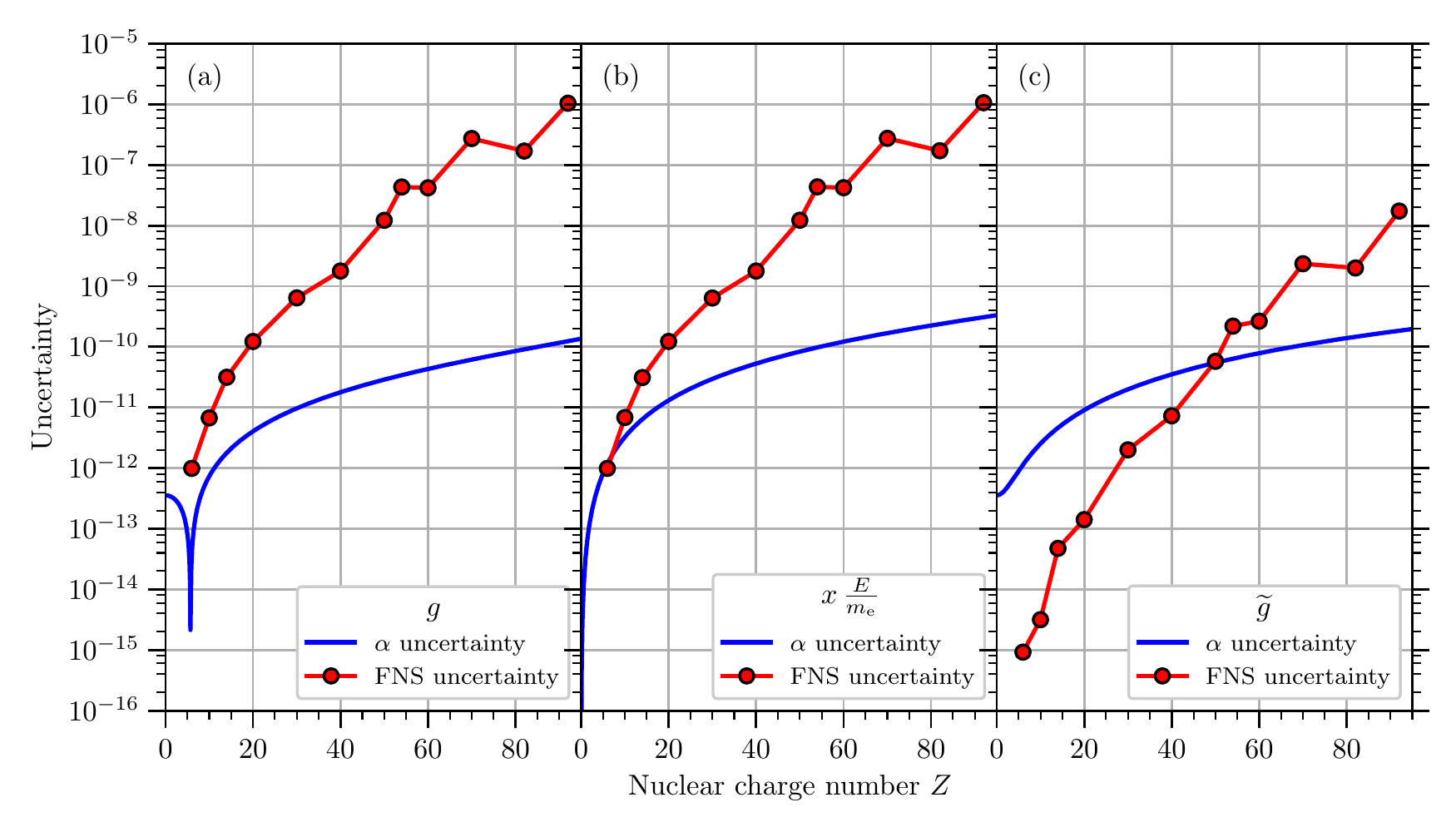}
  \vspace{-6mm}
  \caption{
    Comparison of the uncertainties $\delta{g}$ due to the absolute uncertainty
    of the fine-structure constant, $\delta\alpha = 1.1 \cdot 10^{-12}$~\cite{Tiesinga2020} (blue),
    and due to the uncertainty of the finite nuclear size effect (red).
    The comparisons are shown for the ground-state
    $g$~factor of H-like ions in (a), for the weighted dimensionless
    energy $x \cdot E m_{\mathrm{e}}$ in (b), and the reduced $g$~factor
    $\widetilde{g}$ [defined in Eq.~(\ref{eq:defreduced})] in (c).}
  \label{fig:comparison}
\end{figure*}

\textit{Leading finite nuclear size effect.} --- The leading Dirac contribution to the
$g$~factor in the ground ($1s$) state H-like ions is~\cite{Breit1928}
\begin{equation}
  \label{eq:gintegral}
  g_{\mathrm{D}}^{\mathrm{ext}} = -\frac{8}{3} \int_{0}^{\infty} dr\, r G(r) F(r)\,,
\end{equation}
where $G$ and $F$ denote the upper and lower radial components of the
bound wave function
\begin{equation}
  \label{eq:2}
  \psi_{n \kappa m}(\bm{r})
  = \frac{1}{r}
  \begin{pmatrix} 
    G_{n\kappa}(r) \Omega_{\kappa m}(\bm{r}/r) \\
    i F_{n\kappa}(r) \Omega_{-\kappa m}(\bm{r}/r)
  \end{pmatrix} \,,
\end{equation}
with $\Omega_{\kappa m}(\bm{r}/r)$ denoting spherical spinors. $n$ and $\kappa$ are the principal
and Dirac angular momentum quantum numbers, respectively, and $m$ is the magnetic quantum number.
$\psi$ satisfies Dirac's equation
\begin{equation}
  \label{eq:Dirac}
  \left( \boldsymbol{\alpha}\cdot\mathbf{p} + m_{\rm e} {\beta} + V(r)\right) \psi = E \psi \,,
\end{equation}
with the usual Dirac matrices $\boldsymbol{\alpha}$ and $\beta$, the three-momentum operator $\mathbf{p}$,
and the radial nuclear potential $V(r)$.
For a point-like nucleus with $V(r) = -Z\alpha/r$, the integral in Eq.~(\ref{eq:gintegral}) can be evaluated analytically,
yielding the formula for $g_{\mathrm{D}}$ given above.

For an extended nucleus, we calculate the integral in Eq.~(\ref{eq:gintegral})
numerically, solving the radial Dirac equation using the dual kinetic balance~(DKB)
approach~\cite{Shabaev2004} implemented in quadruple precision.
The leading finite nuclear size~(FNS) contribution to the $g$~factor is
$g_{\mathrm{D}}^{\mathrm{fns}} = g_{\mathrm{D}}^{\mathrm{ext}} - g_{\mathrm{D}}$.
We use first the two-parameter Fermi function as the nuclear charge distribution
and take root mean square~(RMS) nuclear radii from Ref.~\cite{Angeli2013}.  There is an
uncertainty $\delta_{\mathrm{RMS}}g_{\mathrm{D}}^{\mathrm{fns}}$ resulting from the
uncertainties of the RMS radii.  Additionally, in order to estimate the
dependence $\delta_{\mathrm{model}}g_{\mathrm{D}}^{\mathrm{fns}}$ of the FNS effect on the
nuclear model, we also calculate the radial distribution of protons performing
Hartree-Fock-Skyrme nuclear structural calculations~\cite{Colo2013}, and take the difference
of the values obtained with the different distributions.
In Ref.~\cite{Valuev2020} it was observed that proton distributions resulting from
Skyrme forces are in good agreement with distributions measured in electron scattering
experiments. The total uncertainty of the FNS contribution is given as the quadratic sum of
$\delta_{\mathrm{RMS}}g_{\mathrm{D}}^{\mathrm{fns}}$ and $\delta_{\mathrm{model}}g_{\mathrm{D}}^{\mathrm{fns}}$.

Fig.~\ref{fig:comparison}a compares the uncertainty of the $g$~factor --- that of
$g_{\mathrm{D}}$ and the dominant radiative correction, the Schwinger term
$\alpha/\pi$ --- due to $\delta\alpha$, the absolute uncertainty of $\alpha$, and the uncertainty caused by the FNS effect.
Already for low values of $Z$, the uncertainties due to FNS are approx. an order of magnitude larger than those due to
$\delta\alpha$, and the discrepancy grows for heavier elements.

The leading contribution to the $1s$ electron energy assuming a point-like nucleus is given by~\cite{Darwin1928}
  $E_{\mathrm{D}} = m_{\mathrm{e}} \sqrt{1 - (Z\alpha)^{2}} = m_{\mathrm{e}} ( 1 - \frac{1}{2}(Z\alpha)^{2} + \dots)$.
For an extended nucleus, we obtain the ground-state energy
$E_{\mathrm{D}}^{\mathrm{ext}}$ from the numerical solution of the radial Dirac equation.
The calculation of $E_{\mathrm{D}}^{\mathrm{fns}} = E_{\mathrm{D}}^{\mathrm{ext}} - E_{\mathrm{D}}$ and the determination of its
uncertainty is performed similarly as above.

Using the Dirac equation~(\ref{eq:Dirac}) and its radial counterpart,
Eq.~(\ref{eq:gintegral}) for the relativistic $g$ factor may be rewritten as~\cite{Karshenboim2005}
\begin{equation}
  \label{eq:magic}
  g_{\mathrm{D}}^{\mathrm{ext}} = \frac{2}{3}\left( 1+2\langle \beta \rangle \right)
                 = \frac{2}{3}\left( 1+2\frac{\partial E_{\mathrm{D}}^{\mathrm{ext}}}{\partial m_{\rm e}} \right)\,.
\end{equation}
The FNS correction to the energy can be approximated on the one per thousand level as~\cite{Shabaev1993}
$E_{\mathrm{D}}^{\mathrm{fns}}\sim m_{\rm e} (2Z\alpha m_{\rm e} R)^{2\gamma}$
[with $\gamma=\sqrt{1 - (Z\alpha)^{2}}$ and an effective nuclear radius $R$], therefore,
the FNS effects of the $g$ factor and the energy can be accurately related via the
formula~\cite{Karshenboim2005}
\begin{equation}
  \label{eq:Karshenboim}
  g_{\mathrm{D}}^{\mathrm{fns}} \approx \frac{4}{3} \left(2\sqrt{1 - (Z\alpha)^{2}} + 1\right) \frac{E_{\mathrm{D}}^{\mathrm{fns}}}{m_{\mathrm{e}}} \,.
\end{equation}
This motivates the choice of the weighted difference of the $g$ factor and the dimensionless energy in
Eqs.~(\ref{eq:defreduced},\ref{eq:x}): we expect the FNS effect in $\widetilde{g}$ to cancel to a significant degree.
As expected from Eq.~(\ref{eq:Karshenboim}), Fig.~\ref{fig:comparison}b shows that the FNS uncertainties are of comparable magnitude as the
ones for the $g$~factor. Also, as in the case of the $g$~factor, the FNS effect 
causes larger errors than $\delta\alpha$ in the (weighted) energy.

The relevant uncertainties of $\widetilde{g}$ are depicted in Fig.~\ref{fig:comparison}c.
The FNS uncertainties are reduced by up to 2--3 orders of magnitude, and for $Z < 50$, these
uncertainties are smaller than the ones due to $\delta\alpha$, rendering a broad range of elements
ideal for $\alpha$ determination. Furthermore, as compared to the $g$~factor curve, the sensitivity
due to $\delta\alpha$ is generally improved, since
\begin{equation}
  \label{eq:13}
  \frac{\partial}{\partial \alpha}\left.\left(\widetilde{g}_{\mathrm{D}} \right)\right\vert_{{\rm fixed}~x}
  = - 2 \sqrt{1 - (Z\alpha)^{2}} \frac{\partial g_{\mathrm{D}}}{\partial \alpha}
\end{equation}
contains an enhancement factor of magnitude $\approx 2$. Also, in $\widetilde{g}$, the dip in the $g$ factors
$\alpha$-sensitivity around $Z=5$ is removed (see Fig.~\ref{fig:comparison}a). This gives some advantage to the
reduced $g$ factor scheme over those employing weighted differences of $g$ factors in different charge states, since in
those cases the sensitivity to $\delta\alpha$ is slightly reduced in the difference~\cite{ShabaevPRL2006,YerokhinPRL2016}.
In the following, we show that the strong suppression of nuclear effects remains true even
when considering higher-order nuclear contributions.


\textit{QED finite nuclear size effect.} --- QED corrections to the electronic
energy level as well as the $g$~factor arise from one-loop self-energy (SE) and
vacuum polarization (VP) diagrams. The FNS corrections to these terms have been
evaluated e.g. in Ref.~\cite{Yerokhin2011} for the Lamb shift and in
Ref.~\cite{Yerokhin2013} for the $g$~factor. We use the results of
these papers to determine the uncertainty of the QED-FNS effect of the reduced
$g$~factor $\widetilde{g}$ as a quadratic sum of the Lamb shift and the $g$~factor
uncertainties. We find that the uncertainty of the QED-FNS effect
typically raises the FNS uncertainty of the reduced $g$~factor by a factor of 3 or below.
Also, QED FNS has a purely calculational uncertainty, which can be improved further,
and thus the statements of the previous paragraph remain unchanged.


\textit{Nuclear polarization correction.} --- In the above calculations, the protonic charge
distribution was assumed to be that of an isolated, bare nucleus. In an atom, additional small
effects arise from the mutual polarization of protons and electrons.
A simple approximation of this nuclear polarization (NP) correction to energy levels can be found in Ref.~\cite{Flambaum2018},
which can be also extended to the $g$ factor~\cite{Debierre2019}. For the NP correction to the reduced $g$ factor we obtain
the simple analytical formula
\begin{equation}
  \label{eq:analyticalNP}
  \widetilde{g}^{\mathrm{NP}} \approx \frac{32}{3} m_{\rm e}^3 \alpha_{\rm d} (Z \alpha)^4 \frac{\Gamma(2\gamma-2;2Z\alpha R m_{\rm e})}{\Gamma(2\gamma+1)}\,,
\end{equation}
expressed in terms of the radius $R$ of the homogeneously charged sphere model and the dipole nuclear polarizability
$\alpha_{\rm d}$. The latter can be approximated in Migdal's theory~\cite{Levinger1957} as
$\alpha_{\rm d}=\zeta(A)A\alpha \cfrac{R^2}{40 a_{\rm sym}}$, with $a_{\rm sym}=$23~MeV and
$\zeta\left(A\right)=0.76+{2.79}/{A^{1/3}}$. While this simple model yields 
an order-of-magnitude estimate of the effect, it shows a 2-3 digits cancellation of the NP correction in $\widetilde{g}$.

\begin{table}
\begin{center}
\begin{tabular} {r r c l l}
\hline \hline
Ion	\qquad & \qquad $R$ (fm)	& $E_{\rm NP}$ (meV) & $g_{\rm NP}$ & $\widetilde{g}_{\rm NP}$\\
\hline
${}_{10}^{22}$Ne	& 2.9525	& -0.00024	& -2.10[-12]	& -2.39[-13] \\
${}_{14}^{28}$Si	& 3.1224	& -0.00105	& -9.07[-12]	& -8.77[-13] \\
${}_{20}^{40}$Ca	& 3.4776	& -0.00607	& -5.11[-11]	& -3.95[-12] \\
${}_{30}^{64}$Zn	& 3.9283	& -0.0545	& -4.45[-10]	& -2.56[-11] \\
${}_{36}^{84}$Kr	& 4.1884	& -0.144	& -1.15[-9]	& -5.87[-11] \\
${}_{44}^{102}$Ru	& 4.4809	& -0.541	& -4.25[-9]	& -1.66[-10] \\
${}_{48}^{112}$Cd	& 4.5944 	& -0.857	& -6.66[-9]	& -2.34[-10] \\
${}_{60}^{142}$Nd	& 4.9123	& -2.96		& -2.22[-8]	& -5.53[-10] \\
${}_{64}^{158}$Gd	& 5.1569	& -10.4		& -7.69[-8]	& -1.62[-9] \\
${}_{66}^{162}$Dy	& 5.2074	& -12.9		& -9.47[-8]	& -1.86[-9] \\
${}_{70}^{174}$Yb	& 5.3108	& -18.9		& -1.37[-7]	& -2.32[-9] \\
${}_{78}^{196}$Pt	& 5.4307	& -22.6		& -1.57[-7]	& -1.74[-9] \\
${}_{82}^{208}$Pb	& 5.5012	& -28.9		& -1.98[-7]	& -1.54[-9] \\
${}_{92}^{238}$U	& 5.5817	& -196.5	& -1.27[-6]	& -2.09[-9] \\
\hline \hline
\end{tabular} 
\caption{Nuclear polarization corrections to the energy, $g$ factor and reduced $g$ factor. The numbers in brackets indicate powers of 10.
\label{tab:NP}}
\end{center}
\end{table}

A more sophisticated evaluation	 of the NP correction can be performed in the framework of perturbation theory following
Refs.~\cite{Labzowsky1994,Nefiodov1996,Nefiodov2003,Volotka2014}. In this formalism, the electronic reference and intermediate
states are treated relativistically, and nuclear transition data are taken from tabulations. The NP correction to the level
energy reads:
\begin{eqnarray}
\label{eq:ENP}
&&E^{\rm NP} =  - \alpha \sum\limits_{LM} \sum_j B(EL) \nonumber \\
&&\times \int\limits_{-\infty}^{\infty} \frac{d\omega}{2\pi i}\frac{2\omega_L}{\omega^2 - \omega_L^2 +i0}
\frac{|\langle 1s|F_L Y_{LM}|j\rangle|^2}{\epsilon_{1s} - \omega - \epsilon_j(1-i0)}\,,
\end{eqnarray}
where the summation extends over all nuclear excitation energies $\omega_L$ with the reduced electric multipole transition strengths
$B(EL; L \rightarrow 0)$ ($L$ is the multipolarity of a $2^L$-pole transition). The label $j$ denotes intermediate electronic states in the
Dirac spectrum, the $\epsilon_j$ are their unperturbed eigenvalues, and the $Y_{LM}$ denote spherical harmonics.
The radial part is given in the sharp-surface approximation~\cite{Plunien1991},
\begin{eqnarray}
F_L = \left\{
                \begin{array}{ll}
                  \frac{5\sqrt{\pi}}{2R^3}\biggl[1-\biggl(\frac{r}{R}\biggr)^2\biggr]\Theta(R-r)\,, & L=0\,,\\
                  \frac{4\pi}{(2L+1)R^L} \frac{\left({\rm min}(r,R)\right)^L}{\left({\rm max}(r,R)\right)^{L+1}}\,,           & L \geq 1\,.
                \end{array}
              \right.
\end{eqnarray}  
For the NP correction to the $g$ factor, one can write
\begin{eqnarray}
\label{eq:gNP}
g^{\rm NP}  = -m_{\rm e} \frac{\alpha}{m} \sum\limits_{LM} \sum\limits_{j,k} B(EL)
\int\limits_{-\infty}^{\infty} \frac{d\omega}{2\pi i}\frac{2\omega_L}{\omega^2 - \omega_L^2 +i0} \nonumber \\
\times \left[ 2\frac{\langle 1s|F_L Y^*_{LM}|j\rangle
\langle j|F_L Y_{LM}|k\rangle\langle k| \left[\mathbf{r}\times\boldsymbol{\alpha}\right]_z |1s\rangle}
{(\epsilon_{1s} - \omega - \epsilon_j(1-i0))(\epsilon_{1s}  - \epsilon_k(1-i0))} \right. \nonumber \\
\left. + \frac{\langle 1s|F_L Y^*_{LM}|j\rangle \langle j|  \left[\mathbf{r}\times\boldsymbol{\alpha}\right]_z |k\rangle \langle k|F_L Y_{LM}|1s\rangle}
{(\epsilon_{1s} - \omega - \epsilon_j(1-i0))(\epsilon_{1s}  - \omega-\epsilon_k(1-i0))}
\right]\,\,\,
\end{eqnarray}
with $\left[\dots\right]_z$ denoting the $z$ component of a vector. Here, the first summand in the brackets corresponds to the reducible ($k=1s$)
and irreducible ($k\neq 1s$) contributions, and the second to the vertex contribution.

\begin{figure}[t!]
  \centering
  \includegraphics[width=0.8\columnwidth]{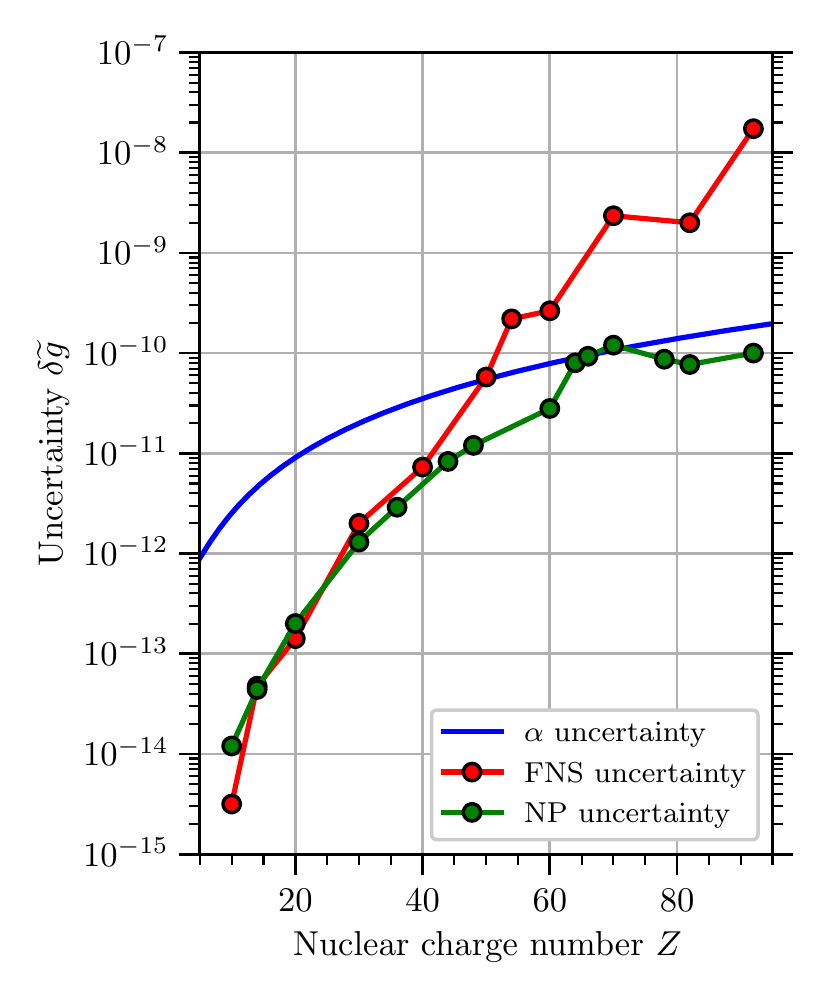}
  \vspace{-6mm}
  \caption{
  Uncertainties of $\widetilde{g}$ due to the nuclear polarization (NP) effect compared to those arising from the leading finite
  nuclear size (FNS) effect, for H-like ions with atomic number~$Z$.
  \label{fig:np}
  }
\end{figure}

The nuclear parameters $\omega_L$ and $B(EL)$ for low-lying nuclear states are taken from Refs.~\cite{22,28,40,64,84,102,112,142,158,162,174,196,208,238}.
The contributions of giant nuclear resonances are estimated by means of phenomenological energy-weighted sum rules \cite{RINKER1978}.
Monopole, dipole, quadrupole and octupole ($L=0$--$3$) low-lying transitions and giant resonances were taken into account.
The spectral summation over the electronic states $j,k$ was performed using the DKB method.
The values obtained for $E^{\rm NP}$ and $g^{\rm NP}$ are in a good agreement with literature values~\cite{Nefiodov1996, Nefiodov2003, Volotka2014} for all available ions.
In Table~\ref{tab:NP}, a significant cancellation of the NP effect can be observed in the reduced $g$~factor.
Additionally, by analyzing the individual contributions from each nuclear transition,
we found that, for the reduced $g$ factor, a detailed knowledge of the nuclear level structures is not needed.
It is sufficient
to take into account only the few strongest transitions with the largest $B(EL)$ to provide reasonably accurate predictions.
Whereas relative uncertainties of $E^{\rm NP}$ and $g^{\rm NP}$ individually reach up to 30-50\%~\cite{Nefiodov1996,Nefiodov2003,Labzowsky1994},
due to effective cancellations between them, we observe that the fractional uncertainty in $\widetilde{g}$ can be
conservatively estimated to be on the few \% level. Assuming a 5\% theoretical uncertainty for $\widetilde{g}^{\rm NP}$,
we find that it is of the same magnitude as the uncertainty of the FNS effect (see Fig.~\ref{fig:np}),
allowing an improved extraction of $\alpha$.

\begin{table}[t!]
\begin{center}
\begin{minipage}{1\linewidth}
\begin{center}
\caption{Numerical values of various contributions to the reduced $g$ factor (\ref{eq:defreduced})
for H-like ${}^{28}\text{Si}^{13+}$. Values of $E$ are taken from Ref.~\cite{VladVladTable} unless otherwise specified
in the reference (Ref.) column. See the text for further details.
\label{tab:Contrib}
} 
\begin{tabular}{llll}
\hline
\hline
Contribution &  & \multicolumn{1}{l}{Value} & Ref.\\
\hline
Dirac    &             & $-1.972\,167\,292\,037\,3(42)$ & \cite{BreitG,CODATA14}\\
One-loop & SE          & $\phantom{-}0.002\,324\,942\,334\,0(59)$ & \cite{VladZOneLoop}\\
         & VP          & $\phantom{-}0.000\,000\,011\,988\,4(72)$ & \cite{VladZOneLoop,LeeDelbrueck}\\
Two-loop &             & $-0.000\,003\,550\,19(60)$ & \cite{CzarneckiEarly,EveryoneTwoLoop,CzarneckiSzafron,UlrichTwoLoop,CzarneckiLetter,VladZVP}\\
$\geq$ Three-loop      & $\left(Z\alpha\right)^0$ & $\phantom{-}0.000\,000\,029\,497\,8$ & \cite{BastianPhD}\\
                       & $\left(Z\alpha\right)^2$ & $\phantom{-}0.000\,000\,000\,051\,5$ & \cite{BastianPhD,CzarneckiEarly}\\
                       & $\left(Z\alpha\right)^{4+}$ & $-0.000\,000\,000\,000\,4(58)$ & \cite{ThreeLoopSlope,ThreeLoopRad}\\
Nucl. recoil & $m_e/M$ & $\phantom{-}0.000\,000\,205\,139\,4(70)$ & \cite{RecoilAllOrders}\\
             &  $\left(m_e/M\right)^{2+}$ & $-0.000\,000\,000\,060\,1(1)$ & \cite{HOMass}\\
             & Rad. recoil & $-0.000\,000\,000\,153(17)$      & \cite{ShabaevReview}\\
FNS          & Leading     & $-0.000\,000\,000\,009\,716(47)$ & \\
NP           &             & $-0.000\,000\,000\,000\,88(5)$   & \\
\hline
Total        &             & $-1.969\,845\,653\,45(60)$       & \\
\hline
\hline
\end{tabular}
\end{center}
\end{minipage}
\end{center}
\end{table}

\textit{Feasibility.} ---
The current status of theory is summarized in Table~\ref{tab:Contrib}, listing the various contributions to the reduced $g$ factor of ${}^{28}\text{Si}^{13+}$.
This numerical example reiterates that nuclear effects do not hinder the extraction of $\alpha$. To this end, one-loop $g$ factor and three-loop
Lamb shift terms, as well as recoil corrections need to be improved by a factor of 1.5--2 at least. As for two-loop diagrams,
the ongoing nonperturbative evaluation~\cite{Sikora2020,VladZVP,Oreshkina2020} of all diagrams needs to be continued, and an evaluation of terms of
order $(\alpha/\pi)(Z\alpha)^6$ in the framework of nonrelativistic QED~\cite{EveryoneTwoLoop,CzarneckiLetter} is desirable. We note that less
substantial theoretical improvements are needed for lighter elements, e.g. for ${}^{12}\text{C}^{5+}$.
As for the experimental prospects: the $g$ factor of HCI can be nowadays measured with relative uncertainties on the level of $10^{-11}$, and further improvement
is possible~\cite{Sturm2019}, allowing a broad range of ions (see Fig.~\ref{fig:comparison}) as candidates.
Fig.~\ref{fig:comparison}b shows that the current $\sim10{}^{-8}$ fractional uncertainty of the total ground-state energy~\cite{Kubicek2014} has to be
decreased by 3-4 orders of magnitude in order to determine $\alpha$ with its present error bar. Recent developments in x-ray spectroscopy of HCI,
i.e. the application of synchrotron and x-ray free electron laser sources~\cite{Rudolph2013,Epp2007,Bernitt2012}, the development of XUV and x-ray frequency combs~\cite{Nauta2017}
as well as advanced laser cooling schemes~\cite{Schmoeger2015,Micke2020} indicate that this goal can be reached and surpassed. Furthermore, the difference
of reduced $g$ factors for two ions with different nuclear charges $Z_1$ and $Z_2$ may also be considered: differential measurements are typically more accurate than absolute
ones, while the corresponding sensitivity to $\alpha$, namely,
$\partial \widetilde{g}(Z_1)/\partial \alpha-\partial \widetilde{g}(Z_2)/\partial \alpha \approx (8/3) \alpha (Z_1^2-Z_2^2)$
is comparable to that of a single-ion measurement.

In summary, the reduced $g$~factor of a simple one-electron ion, i.e. a combination of the bound-electron $g$ factor
and the ground-state energy [given by Eqs.~(\ref{eq:defreduced},\ref{eq:x})],
is put forward as an efficient means for the determination of the fine-structure constant from experimental
data on these atomic quantities. The reduced $g$ factor features a strongly suppressed sensitivity to nuclear effects,
and an enhanced sensitivity to $\alpha$ as compared to the regular $g$ factor. By evaluating and analyzing the
finite nuclear size and nuclear polarization effects on the $g$ factor, the binding energy, and their radiative corrections,
we show that existing and currently developed experimental technology, together with theoretical progress, will allow improving the
uncertainty of $\alpha$ by orders of magnitude in the foreseeable future.

This work is part of and supported by the German Research Foundation (DFG) Collaborative Research Centre ``SFB 1225 (ISOQUANT).''


\begin{thebibliography}{85}%
\makeatletter
\providecommand \@ifxundefined [1]{%
 \@ifx{#1\undefined}
}%
\providecommand \@ifnum [1]{%
 \ifnum #1\expandafter \@firstoftwo
 \else \expandafter \@secondoftwo
 \fi
}%
\providecommand \@ifx [1]{%
 \ifx #1\expandafter \@firstoftwo
 \else \expandafter \@secondoftwo
 \fi
}%
\providecommand \natexlab [1]{#1}%
\providecommand \enquote  [1]{``#1''}%
\providecommand \bibnamefont  [1]{#1}%
\providecommand \bibfnamefont [1]{#1}%
\providecommand \citenamefont [1]{#1}%
\providecommand \href@noop [0]{\@secondoftwo}%
\providecommand \href [0]{\begingroup \@sanitize@url \@href}%
\providecommand \@href[1]{\@@startlink{#1}\@@href}%
\providecommand \@@href[1]{\endgroup#1\@@endlink}%
\providecommand \@sanitize@url [0]{\catcode `\\12\catcode `\$12\catcode
  `\&12\catcode `\#12\catcode `\^12\catcode `\_12\catcode `\%12\relax}%
\providecommand \@@startlink[1]{}%
\providecommand \@@endlink[0]{}%
\providecommand \url  [0]{\begingroup\@sanitize@url \@url }%
\providecommand \@url [1]{\endgroup\@href {#1}{\urlprefix }}%
\providecommand \urlprefix  [0]{URL }%
\providecommand \Eprint [0]{\href }%
\providecommand \doibase [0]{http://dx.doi.org/}%
\providecommand \selectlanguage [0]{\@gobble}%
\providecommand \bibinfo  [0]{\@secondoftwo}%
\providecommand \bibfield  [0]{\@secondoftwo}%
\providecommand \translation [1]{[#1]}%
\providecommand \BibitemOpen [0]{}%
\providecommand \bibitemStop [0]{}%
\providecommand \bibitemNoStop [0]{.\EOS\space}%
\providecommand \EOS [0]{\spacefactor3000\relax}%
\providecommand \BibitemShut  [1]{\csname bibitem#1\endcsname}%
\let\auto@bib@innerbib\@empty
\bibitem [{\citenamefont {Bouchendira}\ \emph {et~al.}(2011)\citenamefont
  {Bouchendira}, \citenamefont {Clad\'e}, \citenamefont {Guellati-Kh\'elifa},
  \citenamefont {Nez},\ and\ \citenamefont {Biraben}}]{Bouchendira2011}%
  \BibitemOpen
  \bibfield  {author} {\bibinfo {author} {\bibfnamefont {R.}~\bibnamefont
  {Bouchendira}}, \bibinfo {author} {\bibfnamefont {P.}~\bibnamefont
  {Clad\'e}}, \bibinfo {author} {\bibfnamefont {S.}~\bibnamefont
  {Guellati-Kh\'elifa}}, \bibinfo {author} {\bibfnamefont {F.}~\bibnamefont
  {Nez}}, \ and\ \bibinfo {author} {\bibfnamefont {F.}~\bibnamefont
  {Biraben}},\ }\href@noop {} {\bibfield  {journal} {\bibinfo  {journal} {Phys.
  Rev. Lett.}\ }\textbf {\bibinfo {volume} {106}},\ \bibinfo {pages} {080801}
  (\bibinfo {year} {2011})}\BibitemShut {NoStop}%
\bibitem [{\citenamefont {Parker}\ \emph {et~al.}(2018)\citenamefont {Parker},
  \citenamefont {Yu}, \citenamefont {Zhong}, \citenamefont {Estey},\ and\
  \citenamefont {M{\"u}ller}}]{Parker2018}%
  \BibitemOpen
  \bibfield  {author} {\bibinfo {author} {\bibfnamefont {R.~H.}\ \bibnamefont
  {Parker}}, \bibinfo {author} {\bibfnamefont {C.}~\bibnamefont {Yu}}, \bibinfo
  {author} {\bibfnamefont {W.}~\bibnamefont {Zhong}}, \bibinfo {author}
  {\bibfnamefont {B.}~\bibnamefont {Estey}}, \ and\ \bibinfo {author}
  {\bibfnamefont {H.}~\bibnamefont {M{\"u}ller}},\ }\href {\doibase
  10.1126/science.aap7706} {\bibfield  {journal} {\bibinfo  {journal}
  {Science}\ }\textbf {\bibinfo {volume} {360}},\ \bibinfo {pages} {191}
  (\bibinfo {year} {2018})}\BibitemShut {NoStop}%
\bibitem [{\citenamefont {Hanneke}\ \emph {et~al.}(2008)\citenamefont
  {Hanneke}, \citenamefont {Fogwell},\ and\ \citenamefont
  {Gabrielse}}]{Hanneke2008}%
  \BibitemOpen
  \bibfield  {author} {\bibinfo {author} {\bibfnamefont {D.}~\bibnamefont
  {Hanneke}}, \bibinfo {author} {\bibfnamefont {S.}~\bibnamefont {Fogwell}}, \
  and\ \bibinfo {author} {\bibfnamefont {G.}~\bibnamefont {Gabrielse}},\
  }\href@noop {} {\bibfield  {journal} {\bibinfo  {journal} {Phys. Rev. Lett.}\
  }\textbf {\bibinfo {volume} {100}},\ \bibinfo {eid} {120801} (\bibinfo {year}
  {2008})}\BibitemShut {NoStop}%
\bibitem [{\citenamefont {Mohr}\ \emph {et~al.}(2012)\citenamefont {Mohr},
  \citenamefont {Taylor},\ and\ \citenamefont {Newell}}]{Mohr2012}%
  \BibitemOpen
  \bibfield  {author} {\bibinfo {author} {\bibfnamefont {P.~J.}\ \bibnamefont
  {Mohr}}, \bibinfo {author} {\bibfnamefont {B.~N.}\ \bibnamefont {Taylor}}, \
  and\ \bibinfo {author} {\bibfnamefont {D.~B.}\ \bibnamefont {Newell}},\
  }\href {\doibase 10.1103/RevModPhys.84.1527} {\bibfield  {journal} {\bibinfo
  {journal} {Rev. Mod. Phys.}\ }\textbf {\bibinfo {volume} {84}},\ \bibinfo
  {pages} {1527} (\bibinfo {year} {2012})}\BibitemShut {NoStop}%
\bibitem [{\citenamefont {Tiesinga}\ \emph {et~al.}(2020)\citenamefont
  {Tiesinga}, \citenamefont {Mohr}, \citenamefont {Newell},\ and\ \citenamefont
  {Taylor}}]{Tiesinga2020}%
  \BibitemOpen
  \bibfield  {author} {\bibinfo {author} {\bibfnamefont {E.}~\bibnamefont
  {Tiesinga}}, \bibinfo {author} {\bibfnamefont {P.~J.}\ \bibnamefont {Mohr}},
  \bibinfo {author} {\bibfnamefont {D.~B.}\ \bibnamefont {Newell}}, \ and\
  \bibinfo {author} {\bibfnamefont {B.~N.}\ \bibnamefont {Taylor}},\ }\href
  {http://physics.nist.gov/constants} {\enquote {\bibinfo {title} {{The 2018
  CODATA Recommended Values of the Fundamental Physical Constants}},}\
  }\bibinfo {howpublished} {\url{http://physics.nist.gov/constants}} (\bibinfo
  {year} {2020}),\ \bibinfo {note} {{(Web Version 8.1) Database developed by J.
  Baker, M. Douma, and S. Kotochigova, National Institute of Standards and
  Technology, Gaithersburg, MD 20899}}\BibitemShut {NoStop}%
\bibitem [{\citenamefont {Breit}(1928{\natexlab{a}})}]{Breit1928}%
  \BibitemOpen
  \bibfield  {author} {\bibinfo {author} {\bibfnamefont {G.}~\bibnamefont
  {Breit}},\ }\href {https://doi.org/10.1038/122649a0} {\bibfield  {journal}
  {\bibinfo  {journal} {Nature}\ }\textbf {\bibinfo {volume} {122}},\ \bibinfo
  {pages} {649} (\bibinfo {year} {1928}{\natexlab{a}})}\BibitemShut {NoStop}%
\bibitem [{\citenamefont {Shabaev}\ \emph {et~al.}(2006)\citenamefont
  {Shabaev}, \citenamefont {Glazov}, \citenamefont {Oreshkina}, \citenamefont
  {Volotka}, \citenamefont {Plunien}, \citenamefont {Kluge},\ and\
  \citenamefont {Quint}}]{ShabaevPRL2006}%
  \BibitemOpen
  \bibfield  {author} {\bibinfo {author} {\bibfnamefont {V.~M.}\ \bibnamefont
  {Shabaev}}, \bibinfo {author} {\bibfnamefont {D.~A.}\ \bibnamefont {Glazov}},
  \bibinfo {author} {\bibfnamefont {N.~S.}\ \bibnamefont {Oreshkina}}, \bibinfo
  {author} {\bibfnamefont {A.~V.}\ \bibnamefont {Volotka}}, \bibinfo {author}
  {\bibfnamefont {G.}~\bibnamefont {Plunien}}, \bibinfo {author} {\bibfnamefont
  {H.-J.}\ \bibnamefont {Kluge}}, \ and\ \bibinfo {author} {\bibfnamefont
  {W.}~\bibnamefont {Quint}},\ }\href@noop {} {\bibfield  {journal} {\bibinfo
  {journal} {Phys. Rev. Lett.}\ }\textbf {\bibinfo {volume} {96}},\ \bibinfo
  {pages} {253002} (\bibinfo {year} {2006})}\BibitemShut {NoStop}%
\bibitem [{\citenamefont {Yerokhin}\ \emph {et~al.}(2016)\citenamefont
  {Yerokhin}, \citenamefont {Berseneva}, \citenamefont {Harman}, \citenamefont
  {Tupitsyn},\ and\ \citenamefont {Keitel}}]{YerokhinPRL2016}%
  \BibitemOpen
  \bibfield  {author} {\bibinfo {author} {\bibfnamefont {V.~A.}\ \bibnamefont
  {Yerokhin}}, \bibinfo {author} {\bibfnamefont {E.}~\bibnamefont {Berseneva}},
  \bibinfo {author} {\bibfnamefont {Z.}~\bibnamefont {Harman}}, \bibinfo
  {author} {\bibfnamefont {I.~I.}\ \bibnamefont {Tupitsyn}}, \ and\ \bibinfo
  {author} {\bibfnamefont {C.~H.}\ \bibnamefont {Keitel}},\ }\href {\doibase
  10.1103/PhysRevLett.116.100801} {\bibfield  {journal} {\bibinfo  {journal}
  {Phys. Rev. Lett.}\ }\textbf {\bibinfo {volume} {116}},\ \bibinfo {pages}
  {100801} (\bibinfo {year} {2016})}\BibitemShut {NoStop}%
\bibitem [{\citenamefont {Glazov}\ \emph {et~al.}(2019)\citenamefont {Glazov},
  \citenamefont {K\"ohler-Langes}, \citenamefont {Volotka}, \citenamefont
  {Blaum}, \citenamefont {Hei\ss{}e}, \citenamefont {Plunien}, \citenamefont
  {Quint}, \citenamefont {Rau}, \citenamefont {Shabaev}, \citenamefont
  {Sturm},\ and\ \citenamefont {Werth}}]{Glazov2019}%
  \BibitemOpen
  \bibfield  {author} {\bibinfo {author} {\bibfnamefont {D.~A.}\ \bibnamefont
  {Glazov}}, \bibinfo {author} {\bibfnamefont {F.}~\bibnamefont
  {K\"ohler-Langes}}, \bibinfo {author} {\bibfnamefont {A.~V.}\ \bibnamefont
  {Volotka}}, \bibinfo {author} {\bibfnamefont {K.}~\bibnamefont {Blaum}},
  \bibinfo {author} {\bibfnamefont {F.}~\bibnamefont {Hei\ss{}e}}, \bibinfo
  {author} {\bibfnamefont {G.}~\bibnamefont {Plunien}}, \bibinfo {author}
  {\bibfnamefont {W.}~\bibnamefont {Quint}}, \bibinfo {author} {\bibfnamefont
  {S.}~\bibnamefont {Rau}}, \bibinfo {author} {\bibfnamefont {V.~M.}\
  \bibnamefont {Shabaev}}, \bibinfo {author} {\bibfnamefont {S.}~\bibnamefont
  {Sturm}}, \ and\ \bibinfo {author} {\bibfnamefont {G.}~\bibnamefont
  {Werth}},\ }\href {\doibase 10.1103/PhysRevLett.123.173001} {\bibfield
  {journal} {\bibinfo  {journal} {Phys. Rev. Lett.}\ }\textbf {\bibinfo
  {volume} {123}},\ \bibinfo {pages} {173001} (\bibinfo {year}
  {2019})}\BibitemShut {NoStop}%
\bibitem [{\citenamefont {Arapoglou}\ \emph {et~al.}(2019)\citenamefont
  {Arapoglou}, \citenamefont {Egl}, \citenamefont {H{\"o}cker}, \citenamefont
  {Sailer}, \citenamefont {Tu}, \citenamefont {Weigel}, \citenamefont {Wolf},
  \citenamefont {Cakir}, \citenamefont {Yerokhin}, \citenamefont {Oreshkina},
  \citenamefont {Agababaev}, \citenamefont {Volotka}, \citenamefont {Zinenko},
  \citenamefont {Glazov}, \citenamefont {Harman}, \citenamefont {Keitel},
  \citenamefont {Sturm},\ and\ \citenamefont {Blaum}}]{Arapoglou2019}%
  \BibitemOpen
  \bibfield  {author} {\bibinfo {author} {\bibfnamefont {I.}~\bibnamefont
  {Arapoglou}}, \bibinfo {author} {\bibfnamefont {A.}~\bibnamefont {Egl}},
  \bibinfo {author} {\bibfnamefont {M.}~\bibnamefont {H{\"o}cker}}, \bibinfo
  {author} {\bibfnamefont {T.}~\bibnamefont {Sailer}}, \bibinfo {author}
  {\bibfnamefont {B.}~\bibnamefont {Tu}}, \bibinfo {author} {\bibfnamefont
  {A.}~\bibnamefont {Weigel}}, \bibinfo {author} {\bibfnamefont
  {R.}~\bibnamefont {Wolf}}, \bibinfo {author} {\bibfnamefont {H.}~\bibnamefont
  {Cakir}}, \bibinfo {author} {\bibfnamefont {V.~A.}\ \bibnamefont {Yerokhin}},
  \bibinfo {author} {\bibfnamefont {N.~S.}\ \bibnamefont {Oreshkina}}, \bibinfo
  {author} {\bibfnamefont {V.~A.}\ \bibnamefont {Agababaev}}, \bibinfo {author}
  {\bibfnamefont {A.~V.}\ \bibnamefont {Volotka}}, \bibinfo {author}
  {\bibfnamefont {D.~V.}\ \bibnamefont {Zinenko}}, \bibinfo {author}
  {\bibfnamefont {D.~A.}\ \bibnamefont {Glazov}}, \bibinfo {author}
  {\bibfnamefont {Z.}~\bibnamefont {Harman}}, \bibinfo {author} {\bibfnamefont
  {C.~H.}\ \bibnamefont {Keitel}}, \bibinfo {author} {\bibfnamefont
  {S.}~\bibnamefont {Sturm}}, \ and\ \bibinfo {author} {\bibfnamefont
  {K.}~\bibnamefont {Blaum}},\ }\href {\doibase 10.1103/PhysRevLett.122.253001}
  {\bibfield  {journal} {\bibinfo  {journal} {Phys. Rev. Lett.}\ }\textbf
  {\bibinfo {volume} {122}},\ \bibinfo {pages} {253001} (\bibinfo {year}
  {2019})}\BibitemShut {NoStop}%
\bibitem [{\citenamefont {K{\"o}hler}\ \emph {et~al.}(2016)\citenamefont
  {K{\"o}hler}, \citenamefont {Blaum}, \citenamefont {Block}, \citenamefont
  {Chenmarev}, \citenamefont {Eliseev}, \citenamefont {Glazov}, \citenamefont
  {Goncharov}, \citenamefont {Hou}, \citenamefont {Kracke}, \citenamefont
  {Nesterenko}, \citenamefont {Novikov}, \citenamefont {Quint}, \citenamefont
  {{Minaya Ramirez}}, \citenamefont {Shabaev}, \citenamefont {Sturm},
  \citenamefont {Volotka},\ and\ \citenamefont {Werth}}]{Koehler2016}%
  \BibitemOpen
  \bibfield  {author} {\bibinfo {author} {\bibfnamefont {F.}~\bibnamefont
  {K{\"o}hler}}, \bibinfo {author} {\bibfnamefont {K.}~\bibnamefont {Blaum}},
  \bibinfo {author} {\bibfnamefont {M.}~\bibnamefont {Block}}, \bibinfo
  {author} {\bibfnamefont {S.}~\bibnamefont {Chenmarev}}, \bibinfo {author}
  {\bibfnamefont {S.}~\bibnamefont {Eliseev}}, \bibinfo {author} {\bibfnamefont
  {D.~A.}\ \bibnamefont {Glazov}}, \bibinfo {author} {\bibfnamefont
  {M.}~\bibnamefont {Goncharov}}, \bibinfo {author} {\bibfnamefont
  {J.}~\bibnamefont {Hou}}, \bibinfo {author} {\bibfnamefont {A.}~\bibnamefont
  {Kracke}}, \bibinfo {author} {\bibfnamefont {D.~A.}\ \bibnamefont
  {Nesterenko}}, \bibinfo {author} {\bibfnamefont {Y.~N.}\ \bibnamefont
  {Novikov}}, \bibinfo {author} {\bibfnamefont {W.}~\bibnamefont {Quint}},
  \bibinfo {author} {\bibfnamefont {E.}~\bibnamefont {{Minaya Ramirez}}},
  \bibinfo {author} {\bibfnamefont {V.~M.}\ \bibnamefont {Shabaev}}, \bibinfo
  {author} {\bibfnamefont {S.}~\bibnamefont {Sturm}}, \bibinfo {author}
  {\bibfnamefont {A.~V.}\ \bibnamefont {Volotka}}, \ and\ \bibinfo {author}
  {\bibfnamefont {G.}~\bibnamefont {Werth}},\ }\href@noop {} {\bibfield
  {journal} {\bibinfo  {journal} {Nat. Commun.}\ }\textbf {\bibinfo {volume}
  {7}},\ \bibinfo {pages} {1} (\bibinfo {year} {2016})}\BibitemShut {NoStop}%
\bibitem [{\citenamefont {Wagner}\ \emph {et~al.}(2013)\citenamefont {Wagner},
  \citenamefont {Sturm}, \citenamefont {K\"ohler}, \citenamefont {Glazov},
  \citenamefont {Volotka}, \citenamefont {Plunien}, \citenamefont {Quint},
  \citenamefont {Werth}, \citenamefont {Shabaev},\ and\ \citenamefont
  {Blaum}}]{Wagner2013}%
  \BibitemOpen
  \bibfield  {author} {\bibinfo {author} {\bibfnamefont {A.}~\bibnamefont
  {Wagner}}, \bibinfo {author} {\bibfnamefont {S.}~\bibnamefont {Sturm}},
  \bibinfo {author} {\bibfnamefont {F.}~\bibnamefont {K\"ohler}}, \bibinfo
  {author} {\bibfnamefont {D.~A.}\ \bibnamefont {Glazov}}, \bibinfo {author}
  {\bibfnamefont {A.~V.}\ \bibnamefont {Volotka}}, \bibinfo {author}
  {\bibfnamefont {G.}~\bibnamefont {Plunien}}, \bibinfo {author} {\bibfnamefont
  {W.}~\bibnamefont {Quint}}, \bibinfo {author} {\bibfnamefont
  {G.}~\bibnamefont {Werth}}, \bibinfo {author} {\bibfnamefont {V.~M.}\
  \bibnamefont {Shabaev}}, \ and\ \bibinfo {author} {\bibfnamefont
  {K.}~\bibnamefont {Blaum}},\ }\href {\doibase 10.1103/PhysRevLett.110.033003}
  {\bibfield  {journal} {\bibinfo  {journal} {Phys. Rev. Lett.}\ }\textbf
  {\bibinfo {volume} {110}},\ \bibinfo {pages} {033003} (\bibinfo {year}
  {2013})}\BibitemShut {NoStop}%
\bibitem [{\citenamefont {Yerokhin}\ \emph {et~al.}(2017)\citenamefont
  {Yerokhin}, \citenamefont {Pachucki}, \citenamefont {Puchalski},
  \citenamefont {Harman},\ and\ \citenamefont {Keitel}}]{Yerokhin2017}%
  \BibitemOpen
  \bibfield  {author} {\bibinfo {author} {\bibfnamefont {V.~A.}\ \bibnamefont
  {Yerokhin}}, \bibinfo {author} {\bibfnamefont {K.}~\bibnamefont {Pachucki}},
  \bibinfo {author} {\bibfnamefont {M.}~\bibnamefont {Puchalski}}, \bibinfo
  {author} {\bibfnamefont {Z.}~\bibnamefont {Harman}}, \ and\ \bibinfo {author}
  {\bibfnamefont {C.~H.}\ \bibnamefont {Keitel}},\ }\href {\doibase
  10.1103/PhysRevA.95.062511} {\bibfield  {journal} {\bibinfo  {journal} {Phys.
  Rev. A}\ }\textbf {\bibinfo {volume} {95}},\ \bibinfo {pages} {062511}
  (\bibinfo {year} {2017})}\BibitemShut {NoStop}%
\bibitem [{\citenamefont {Cakir}\ \emph {et~al.}(2020)\citenamefont {Cakir},
  \citenamefont {Yerokhin}, \citenamefont {Oreshkina}, \citenamefont {Sikora},
  \citenamefont {Tupitsyn}, \citenamefont {Keitel},\ and\ \citenamefont
  {Harman}}]{Cakir2020}%
  \BibitemOpen
  \bibfield  {author} {\bibinfo {author} {\bibfnamefont {H.}~\bibnamefont
  {Cakir}}, \bibinfo {author} {\bibfnamefont {V.~A.}\ \bibnamefont {Yerokhin}},
  \bibinfo {author} {\bibfnamefont {N.~S.}\ \bibnamefont {Oreshkina}}, \bibinfo
  {author} {\bibfnamefont {B.}~\bibnamefont {Sikora}}, \bibinfo {author}
  {\bibfnamefont {I.~I.}\ \bibnamefont {Tupitsyn}}, \bibinfo {author}
  {\bibfnamefont {C.~H.}\ \bibnamefont {Keitel}}, \ and\ \bibinfo {author}
  {\bibfnamefont {Z.}~\bibnamefont {Harman}},\ }\href@noop {} {\bibfield
  {journal} {\bibinfo  {journal} {Phys. Rev. A}\ } (\bibinfo {year} {2020})},\
  \bibinfo {note} {{in print}}\BibitemShut {NoStop}%
\bibitem [{\citenamefont {Karshenboim}\ \emph {et~al.}(2005)\citenamefont
  {Karshenboim}, \citenamefont {Lee},\ and\ \citenamefont
  {Milstein}}]{Karshenboim2005}%
  \BibitemOpen
  \bibfield  {author} {\bibinfo {author} {\bibfnamefont {S.~G.}\ \bibnamefont
  {Karshenboim}}, \bibinfo {author} {\bibfnamefont {R.~N.}\ \bibnamefont
  {Lee}}, \ and\ \bibinfo {author} {\bibfnamefont {A.~I.}\ \bibnamefont
  {Milstein}},\ }\href {\doibase 10.1103/PhysRevA.72.042101} {\bibfield
  {journal} {\bibinfo  {journal} {Phys. Rev. A}\ }\textbf {\bibinfo {volume}
  {72}},\ \bibinfo {pages} {042101} (\bibinfo {year} {2005})}\BibitemShut
  {NoStop}%
\bibitem [{\citenamefont {Sturm}\ \emph {et~al.}(2014)\citenamefont {Sturm},
  \citenamefont {K{\"o}hler}, \citenamefont {Zatorski}, \citenamefont {Wagner},
  \citenamefont {Harman}, \citenamefont {Werth}, \citenamefont {Quint},
  \citenamefont {Keitel},\ and\ \citenamefont {Blaum}}]{Sturm2014}%
  \BibitemOpen
  \bibfield  {author} {\bibinfo {author} {\bibfnamefont {S.}~\bibnamefont
  {Sturm}}, \bibinfo {author} {\bibfnamefont {F.}~\bibnamefont {K{\"o}hler}},
  \bibinfo {author} {\bibfnamefont {J.}~\bibnamefont {Zatorski}}, \bibinfo
  {author} {\bibfnamefont {A.}~\bibnamefont {Wagner}}, \bibinfo {author}
  {\bibfnamefont {Z.}~\bibnamefont {Harman}}, \bibinfo {author} {\bibfnamefont
  {G.}~\bibnamefont {Werth}}, \bibinfo {author} {\bibfnamefont
  {W.}~\bibnamefont {Quint}}, \bibinfo {author} {\bibfnamefont {C.~H.}\
  \bibnamefont {Keitel}}, \ and\ \bibinfo {author} {\bibfnamefont
  {K.}~\bibnamefont {Blaum}},\ }\href {\doibase doi:10.1038/nature13026}
  {\bibfield  {journal} {\bibinfo  {journal} {Nature}\ }\textbf {\bibinfo
  {volume} {506}},\ \bibinfo {pages} {467–470} (\bibinfo {year}
  {2014})}\BibitemShut {NoStop}%
\bibitem [{\citenamefont {Sturm}\ \emph
  {et~al.}(2011{\natexlab{a}})\citenamefont {Sturm}, \citenamefont {Wagner},
  \citenamefont {Schabinger}, \citenamefont {Zatorski}, \citenamefont {Harman},
  \citenamefont {Quint}, \citenamefont {Werth}, \citenamefont {Keitel},\ and\
  \citenamefont {Blaum}}]{Sturm2011}%
  \BibitemOpen
  \bibfield  {author} {\bibinfo {author} {\bibfnamefont {S.}~\bibnamefont
  {Sturm}}, \bibinfo {author} {\bibfnamefont {A.}~\bibnamefont {Wagner}},
  \bibinfo {author} {\bibfnamefont {B.}~\bibnamefont {Schabinger}}, \bibinfo
  {author} {\bibfnamefont {J.}~\bibnamefont {Zatorski}}, \bibinfo {author}
  {\bibfnamefont {Z.}~\bibnamefont {Harman}}, \bibinfo {author} {\bibfnamefont
  {W.}~\bibnamefont {Quint}}, \bibinfo {author} {\bibfnamefont
  {G.}~\bibnamefont {Werth}}, \bibinfo {author} {\bibfnamefont {C.~H.}\
  \bibnamefont {Keitel}}, \ and\ \bibinfo {author} {\bibfnamefont
  {K.}~\bibnamefont {Blaum}},\ }\href@noop {} {\bibfield  {journal} {\bibinfo
  {journal} {Phys. Rev. Lett.}\ }\textbf {\bibinfo {volume} {107}},\ \bibinfo
  {pages} {023002} (\bibinfo {year} {2011}{\natexlab{a}})}\BibitemShut
  {NoStop}%
\bibitem [{\citenamefont {Sturm}\ \emph
  {et~al.}(2011{\natexlab{b}})\citenamefont {Sturm}, \citenamefont {Wagner},
  \citenamefont {Schabinger},\ and\ \citenamefont {Blaum}}]{Sturm2011b}%
  \BibitemOpen
  \bibfield  {author} {\bibinfo {author} {\bibfnamefont {S.}~\bibnamefont
  {Sturm}}, \bibinfo {author} {\bibfnamefont {A.}~\bibnamefont {Wagner}},
  \bibinfo {author} {\bibfnamefont {B.}~\bibnamefont {Schabinger}}, \ and\
  \bibinfo {author} {\bibfnamefont {K.}~\bibnamefont {Blaum}},\ }\href@noop {}
  {\bibfield  {journal} {\bibinfo  {journal} {Phys. Rev. Lett.}\ }\textbf
  {\bibinfo {volume} {107}},\ \bibinfo {pages} {143003} (\bibinfo {year}
  {2011}{\natexlab{b}})}\BibitemShut {NoStop}%
\bibitem [{\citenamefont {Decaux}\ \emph {et~al.}(1997)\citenamefont {Decaux},
  \citenamefont {Beiersdorfer}, \citenamefont {Kahn},\ and\ \citenamefont
  {Jacobs}}]{Decaux1997}%
  \BibitemOpen
  \bibfield  {author} {\bibinfo {author} {\bibfnamefont {V.}~\bibnamefont
  {Decaux}}, \bibinfo {author} {\bibfnamefont {P.}~\bibnamefont
  {Beiersdorfer}}, \bibinfo {author} {\bibfnamefont {S.~M.}\ \bibnamefont
  {Kahn}}, \ and\ \bibinfo {author} {\bibfnamefont {V.~L.}\ \bibnamefont
  {Jacobs}},\ }\href {\doibase 10.1086/304169} {\bibfield  {journal} {\bibinfo
  {journal} {The Astrophysical Journal}\ }\textbf {\bibinfo {volume} {482}},\
  \bibinfo {pages} {1076} (\bibinfo {year} {1997})}\BibitemShut {NoStop}%
\bibitem [{\citenamefont {Beiersdorfer}\ \emph {et~al.}(1989)\citenamefont
  {Beiersdorfer}, \citenamefont {Bitter}, \citenamefont {von Goeler},\ and\
  \citenamefont {Hill}}]{Beiersdorfer1989}%
  \BibitemOpen
  \bibfield  {author} {\bibinfo {author} {\bibfnamefont {P.}~\bibnamefont
  {Beiersdorfer}}, \bibinfo {author} {\bibfnamefont {M.}~\bibnamefont
  {Bitter}}, \bibinfo {author} {\bibfnamefont {S.}~\bibnamefont {von Goeler}},
  \ and\ \bibinfo {author} {\bibfnamefont {K.~W.}\ \bibnamefont {Hill}},\
  }\href {\doibase 10.1103/PhysRevA.40.150} {\bibfield  {journal} {\bibinfo
  {journal} {Phys. Rev. A}\ }\textbf {\bibinfo {volume} {40}},\ \bibinfo
  {pages} {150} (\bibinfo {year} {1989})}\BibitemShut {NoStop}%
\bibitem [{\citenamefont {Tavernier}\ \emph {et~al.}(1985)\citenamefont
  {Tavernier}, \citenamefont {Briand}, \citenamefont {Indelicato},
  \citenamefont {Liesen},\ and\ \citenamefont {Richard}}]{Tavernier1985}%
  \BibitemOpen
  \bibfield  {author} {\bibinfo {author} {\bibfnamefont {M.}~\bibnamefont
  {Tavernier}}, \bibinfo {author} {\bibfnamefont {J.~P.}\ \bibnamefont
  {Briand}}, \bibinfo {author} {\bibfnamefont {P.}~\bibnamefont {Indelicato}},
  \bibinfo {author} {\bibfnamefont {D.}~\bibnamefont {Liesen}}, \ and\ \bibinfo
  {author} {\bibfnamefont {P.}~\bibnamefont {Richard}},\ }\href {\doibase
  10.1088/0022-3700/18/11/004} {\bibfield  {journal} {\bibinfo  {journal} {J.
  Phys. B}\ }\textbf {\bibinfo {volume} {18}},\ \bibinfo {pages} {L327}
  (\bibinfo {year} {1985})}\BibitemShut {NoStop}%
\bibitem [{\citenamefont {Marmar}\ \emph {et~al.}(1986)\citenamefont {Marmar},
  \citenamefont {Rice}, \citenamefont {Kallne}, \citenamefont {Kallne},\ and\
  \citenamefont {{LaVilla}}}]{Marmar1986}%
  \BibitemOpen
  \bibfield  {author} {\bibinfo {author} {\bibfnamefont {E.~S.}\ \bibnamefont
  {Marmar}}, \bibinfo {author} {\bibfnamefont {J.~E.}\ \bibnamefont {Rice}},
  \bibinfo {author} {\bibfnamefont {K.}~\bibnamefont {Kallne}}, \bibinfo
  {author} {\bibfnamefont {J.}~\bibnamefont {Kallne}}, \ and\ \bibinfo {author}
  {\bibfnamefont {R.~E.}\ \bibnamefont {{LaVilla}}},\ }\href {\doibase
  10.1103/PhysRevA.33.774} {\bibfield  {journal} {\bibinfo  {journal} {Phys.
  Rev. A}\ }\textbf {\bibinfo {volume} {33}},\ \bibinfo {pages} {774} (\bibinfo
  {year} {1986})}\BibitemShut {NoStop}%
\bibitem [{\citenamefont {Briand}\ \emph {et~al.}(1984)\citenamefont {Briand},
  \citenamefont {Tavernier}, \citenamefont {Marrus},\ and\ \citenamefont
  {Desclaux}}]{Briand1984}%
  \BibitemOpen
  \bibfield  {author} {\bibinfo {author} {\bibfnamefont {J.~P.}\ \bibnamefont
  {Briand}}, \bibinfo {author} {\bibfnamefont {M.}~\bibnamefont {Tavernier}},
  \bibinfo {author} {\bibfnamefont {R.}~\bibnamefont {Marrus}}, \ and\ \bibinfo
  {author} {\bibfnamefont {J.~P.}\ \bibnamefont {Desclaux}},\ }\href {\doibase
  10.1103/PhysRevA.29.3143} {\bibfield  {journal} {\bibinfo  {journal} {Phys.
  Rev. A}\ }\textbf {\bibinfo {volume} {29}},\ \bibinfo {pages} {3143}
  (\bibinfo {year} {1984})}\BibitemShut {NoStop}%
\bibitem [{\citenamefont {Deslattes}\ \emph {et~al.}(1984)\citenamefont
  {Deslattes}, \citenamefont {Beyer},\ and\ \citenamefont
  {Folkmann}}]{Deslattes1984}%
  \BibitemOpen
  \bibfield  {author} {\bibinfo {author} {\bibfnamefont {R.~D.}\ \bibnamefont
  {Deslattes}}, \bibinfo {author} {\bibfnamefont {H.~F.}\ \bibnamefont
  {Beyer}}, \ and\ \bibinfo {author} {\bibfnamefont {F.}~\bibnamefont
  {Folkmann}},\ }\href {\doibase 10.1088/0022-3700/17/21/001} {\bibfield
  {journal} {\bibinfo  {journal} {J. Phys. B}\ }\textbf {\bibinfo {volume}
  {17}},\ \bibinfo {pages} {L689} (\bibinfo {year} {1984})}\BibitemShut
  {NoStop}%
\bibitem [{\citenamefont {Kubi\ifmmode~\check{c}\else \v{c}\fi{}ek}\ \emph
  {et~al.}(2014)\citenamefont {Kubi\ifmmode~\check{c}\else \v{c}\fi{}ek},
  \citenamefont {Mokler}, \citenamefont {M\"ackel}, \citenamefont {Ullrich},\
  and\ \citenamefont {{Crespo L\'opez-Urrutia}}}]{Kubicek2014}%
  \BibitemOpen
  \bibfield  {author} {\bibinfo {author} {\bibfnamefont {K.}~\bibnamefont
  {Kubi\ifmmode~\check{c}\else \v{c}\fi{}ek}}, \bibinfo {author} {\bibfnamefont
  {P.~H.}\ \bibnamefont {Mokler}}, \bibinfo {author} {\bibfnamefont
  {V.}~\bibnamefont {M\"ackel}}, \bibinfo {author} {\bibfnamefont
  {J.}~\bibnamefont {Ullrich}}, \ and\ \bibinfo {author} {\bibfnamefont
  {J.~R.}\ \bibnamefont {{Crespo L\'opez-Urrutia}}},\ }\href {\doibase
  10.1103/PhysRevA.90.032508} {\bibfield  {journal} {\bibinfo  {journal} {Phys.
  Rev. A}\ }\textbf {\bibinfo {volume} {90}},\ \bibinfo {pages} {032508}
  (\bibinfo {year} {2014})}\BibitemShut {NoStop}%
\bibitem [{\citenamefont {Bruhns}\ \emph {et~al.}(2007)\citenamefont {Bruhns},
  \citenamefont {Braun}, \citenamefont {Kubi\ifmmode~\check{c}\else
  \v{c}\fi{}ek}, \citenamefont {Crespo L\'opez-Urrutia},\ and\ \citenamefont
  {Ullrich}}]{Bruhns2007}%
  \BibitemOpen
  \bibfield  {author} {\bibinfo {author} {\bibfnamefont {H.}~\bibnamefont
  {Bruhns}}, \bibinfo {author} {\bibfnamefont {J.}~\bibnamefont {Braun}},
  \bibinfo {author} {\bibfnamefont {K.}~\bibnamefont
  {Kubi\ifmmode~\check{c}\else \v{c}\fi{}ek}}, \bibinfo {author} {\bibfnamefont
  {J.~R.}\ \bibnamefont {Crespo L\'opez-Urrutia}}, \ and\ \bibinfo {author}
  {\bibfnamefont {J.}~\bibnamefont {Ullrich}},\ }\href {\doibase
  10.1103/PhysRevLett.99.113001} {\bibfield  {journal} {\bibinfo  {journal}
  {Phys. Rev. Lett.}\ }\textbf {\bibinfo {volume} {99}},\ \bibinfo {pages}
  {113001} (\bibinfo {year} {2007})}\BibitemShut {NoStop}%
\bibitem [{\citenamefont {Rudolph}\ \emph {et~al.}(2013)\citenamefont
  {Rudolph}, \citenamefont {Bernitt}, \citenamefont {Epp}, \citenamefont
  {Steinbr\"ugge}, \citenamefont {Beilmann}, \citenamefont {Brown},
  \citenamefont {Eberle}, \citenamefont {Graf}, \citenamefont {Harman},
  \citenamefont {Hell}, \citenamefont {Leutenegger}, \citenamefont {M\"uller},
  \citenamefont {Schlage}, \citenamefont {Wille}, \citenamefont
  {Yava\ifmmode~\mbox{\c{s}}\else \c{s}\fi{}}, \citenamefont {Ullrich},\ and\
  \citenamefont {Crespo L\'opez-Urrutia}}]{Rudolph2013}%
  \BibitemOpen
  \bibfield  {author} {\bibinfo {author} {\bibfnamefont {J.~K.}\ \bibnamefont
  {Rudolph}}, \bibinfo {author} {\bibfnamefont {S.}~\bibnamefont {Bernitt}},
  \bibinfo {author} {\bibfnamefont {S.~W.}\ \bibnamefont {Epp}}, \bibinfo
  {author} {\bibfnamefont {R.}~\bibnamefont {Steinbr\"ugge}}, \bibinfo {author}
  {\bibfnamefont {C.}~\bibnamefont {Beilmann}}, \bibinfo {author}
  {\bibfnamefont {G.~V.}\ \bibnamefont {Brown}}, \bibinfo {author}
  {\bibfnamefont {S.}~\bibnamefont {Eberle}}, \bibinfo {author} {\bibfnamefont
  {A.}~\bibnamefont {Graf}}, \bibinfo {author} {\bibfnamefont {Z.}~\bibnamefont
  {Harman}}, \bibinfo {author} {\bibfnamefont {N.}~\bibnamefont {Hell}},
  \bibinfo {author} {\bibfnamefont {M.}~\bibnamefont {Leutenegger}}, \bibinfo
  {author} {\bibfnamefont {A.}~\bibnamefont {M\"uller}}, \bibinfo {author}
  {\bibfnamefont {K.}~\bibnamefont {Schlage}}, \bibinfo {author} {\bibfnamefont
  {H.-C.}\ \bibnamefont {Wille}}, \bibinfo {author} {\bibfnamefont
  {H.}~\bibnamefont {Yava\ifmmode~\mbox{\c{s}}\else \c{s}\fi{}}}, \bibinfo
  {author} {\bibfnamefont {J.}~\bibnamefont {Ullrich}}, \ and\ \bibinfo
  {author} {\bibfnamefont {J.~R.}\ \bibnamefont {Crespo L\'opez-Urrutia}},\
  }\href {\doibase 10.1103/PhysRevLett.111.103002} {\bibfield  {journal}
  {\bibinfo  {journal} {Phys. Rev. Lett.}\ }\textbf {\bibinfo {volume} {111}},\
  \bibinfo {pages} {103002} (\bibinfo {year} {2013})}\BibitemShut {NoStop}%
\bibitem [{\citenamefont {Epp}\ \emph {et~al.}(2007)\citenamefont {Epp},
  \citenamefont {{Crespo L\'opez-Urrutia}}, \citenamefont {Brenner},
  \citenamefont {M\"ackel}, \citenamefont {Mokler}, \citenamefont {Treusch},
  \citenamefont {Kuhlmann}, \citenamefont {Yurkov}, \citenamefont {Feldhaus},
  \citenamefont {Schneider}, \citenamefont {Wellh\"ofer}, \citenamefont
  {Martins}, \citenamefont {Wurth},\ and\ \citenamefont {Ullrich}}]{Epp2007}%
  \BibitemOpen
  \bibfield  {author} {\bibinfo {author} {\bibfnamefont {S.~W.}\ \bibnamefont
  {Epp}}, \bibinfo {author} {\bibfnamefont {J.~R.}\ \bibnamefont {{Crespo
  L\'opez-Urrutia}}}, \bibinfo {author} {\bibfnamefont {G.}~\bibnamefont
  {Brenner}}, \bibinfo {author} {\bibfnamefont {V.}~\bibnamefont {M\"ackel}},
  \bibinfo {author} {\bibfnamefont {P.~H.}\ \bibnamefont {Mokler}}, \bibinfo
  {author} {\bibfnamefont {R.}~\bibnamefont {Treusch}}, \bibinfo {author}
  {\bibfnamefont {M.}~\bibnamefont {Kuhlmann}}, \bibinfo {author}
  {\bibfnamefont {M.~V.}\ \bibnamefont {Yurkov}}, \bibinfo {author}
  {\bibfnamefont {J.}~\bibnamefont {Feldhaus}}, \bibinfo {author}
  {\bibfnamefont {J.~R.}\ \bibnamefont {Schneider}}, \bibinfo {author}
  {\bibfnamefont {M.}~\bibnamefont {Wellh\"ofer}}, \bibinfo {author}
  {\bibfnamefont {M.}~\bibnamefont {Martins}}, \bibinfo {author} {\bibfnamefont
  {W.}~\bibnamefont {Wurth}}, \ and\ \bibinfo {author} {\bibfnamefont
  {J.}~\bibnamefont {Ullrich}},\ }\href {\doibase
  10.1103/PhysRevLett.98.183001} {\bibfield  {journal} {\bibinfo  {journal}
  {Phys. Rev. Lett.}\ }\textbf {\bibinfo {volume} {98}},\ \bibinfo {pages}
  {183001} (\bibinfo {year} {2007})}\BibitemShut {NoStop}%
\bibitem [{\citenamefont {Bernitt}\ \emph {et~al.}(2012)\citenamefont
  {Bernitt}, \citenamefont {Brown}, \citenamefont {Rudolph}, \citenamefont
  {Steinbr\"ugge}, \citenamefont {Graf}, \citenamefont {Leutenegger},
  \citenamefont {Epp}, \citenamefont {Eberle}, \citenamefont {Kubi\v{c}ek},
  \citenamefont {M\"ackel}, \citenamefont {Simon}, \citenamefont {Tr\"abert},
  \citenamefont {Magee}, \citenamefont {Beilmann}, \citenamefont {Hell},
  \citenamefont {Schippers}, \citenamefont {M\"uller}, \citenamefont {Kahn},
  \citenamefont {Surzhykov}, \citenamefont {Harman}, \citenamefont {Keitel},
  \citenamefont {Clementson}, \citenamefont {Porter}, \citenamefont
  {Schlotter}, \citenamefont {Turner}, \citenamefont {Ullrich}, \citenamefont
  {Beiersdorfer},\ and\ \citenamefont {{Crespo
  L\'opez-Urrutia}}}]{Bernitt2012}%
  \BibitemOpen
  \bibfield  {author} {\bibinfo {author} {\bibfnamefont {S.}~\bibnamefont
  {Bernitt}}, \bibinfo {author} {\bibfnamefont {G.~V.}\ \bibnamefont {Brown}},
  \bibinfo {author} {\bibfnamefont {J.~K.}\ \bibnamefont {Rudolph}}, \bibinfo
  {author} {\bibfnamefont {R.}~\bibnamefont {Steinbr\"ugge}}, \bibinfo {author}
  {\bibfnamefont {A.}~\bibnamefont {Graf}}, \bibinfo {author} {\bibfnamefont
  {M.}~\bibnamefont {Leutenegger}}, \bibinfo {author} {\bibfnamefont {S.~W.}\
  \bibnamefont {Epp}}, \bibinfo {author} {\bibfnamefont {S.}~\bibnamefont
  {Eberle}}, \bibinfo {author} {\bibfnamefont {K.}~\bibnamefont {Kubi\v{c}ek}},
  \bibinfo {author} {\bibfnamefont {V.}~\bibnamefont {M\"ackel}}, \bibinfo
  {author} {\bibfnamefont {M.~C.}\ \bibnamefont {Simon}}, \bibinfo {author}
  {\bibfnamefont {E.}~\bibnamefont {Tr\"abert}}, \bibinfo {author}
  {\bibfnamefont {E.~W.}\ \bibnamefont {Magee}}, \bibinfo {author}
  {\bibfnamefont {C.}~\bibnamefont {Beilmann}}, \bibinfo {author}
  {\bibfnamefont {N.}~\bibnamefont {Hell}}, \bibinfo {author} {\bibfnamefont
  {S.}~\bibnamefont {Schippers}}, \bibinfo {author} {\bibfnamefont
  {A.}~\bibnamefont {M\"uller}}, \bibinfo {author} {\bibfnamefont {S.~M.}\
  \bibnamefont {Kahn}}, \bibinfo {author} {\bibfnamefont {A.}~\bibnamefont
  {Surzhykov}}, \bibinfo {author} {\bibfnamefont {Z.}~\bibnamefont {Harman}},
  \bibinfo {author} {\bibfnamefont {C.~H.}\ \bibnamefont {Keitel}}, \bibinfo
  {author} {\bibfnamefont {J.}~\bibnamefont {Clementson}}, \bibinfo {author}
  {\bibfnamefont {F.~S.}\ \bibnamefont {Porter}}, \bibinfo {author}
  {\bibfnamefont {W.}~\bibnamefont {Schlotter}}, \bibinfo {author}
  {\bibfnamefont {J.~J.}\ \bibnamefont {Turner}}, \bibinfo {author}
  {\bibfnamefont {J.}~\bibnamefont {Ullrich}}, \bibinfo {author} {\bibfnamefont
  {P.}~\bibnamefont {Beiersdorfer}}, \ and\ \bibinfo {author} {\bibfnamefont
  {J.~R.}\ \bibnamefont {{Crespo L\'opez-Urrutia}}},\ }\href@noop {} {\bibfield
   {journal} {\bibinfo  {journal} {Nature}\ }\textbf {\bibinfo {volume}
  {492}},\ \bibinfo {pages} {225} (\bibinfo {year} {2012})}\BibitemShut
  {NoStop}%
\bibitem [{\citenamefont {Gumberidze}\ \emph {et~al.}(2005)\citenamefont
  {Gumberidze}, \citenamefont {St\"ohlker}, \citenamefont {Bana\'s},
  \citenamefont {Beckert}, \citenamefont {Beller}, \citenamefont {Beyer},
  \citenamefont {Bosch}, \citenamefont {Hagmann}, \citenamefont {Kozhuharov},
  \citenamefont {Liesen}, \citenamefont {Nolden}, \citenamefont {Ma},
  \citenamefont {Mokler}, \citenamefont {Steck}, \citenamefont {Sierpowski},\
  and\ \citenamefont {Tashenov}}]{GumberidzePRL2005}%
  \BibitemOpen
  \bibfield  {author} {\bibinfo {author} {\bibfnamefont {A.}~\bibnamefont
  {Gumberidze}}, \bibinfo {author} {\bibfnamefont {T.}~\bibnamefont
  {St\"ohlker}}, \bibinfo {author} {\bibfnamefont {D.}~\bibnamefont {Bana\'s}},
  \bibinfo {author} {\bibfnamefont {K.}~\bibnamefont {Beckert}}, \bibinfo
  {author} {\bibfnamefont {P.}~\bibnamefont {Beller}}, \bibinfo {author}
  {\bibfnamefont {H.~F.}\ \bibnamefont {Beyer}}, \bibinfo {author}
  {\bibfnamefont {F.}~\bibnamefont {Bosch}}, \bibinfo {author} {\bibfnamefont
  {S.}~\bibnamefont {Hagmann}}, \bibinfo {author} {\bibfnamefont
  {C.}~\bibnamefont {Kozhuharov}}, \bibinfo {author} {\bibfnamefont
  {D.}~\bibnamefont {Liesen}}, \bibinfo {author} {\bibfnamefont
  {F.}~\bibnamefont {Nolden}}, \bibinfo {author} {\bibfnamefont
  {X.}~\bibnamefont {Ma}}, \bibinfo {author} {\bibfnamefont {P.~H.}\
  \bibnamefont {Mokler}}, \bibinfo {author} {\bibfnamefont {M.}~\bibnamefont
  {Steck}}, \bibinfo {author} {\bibfnamefont {D.}~\bibnamefont {Sierpowski}}, \
  and\ \bibinfo {author} {\bibfnamefont {S.}~\bibnamefont {Tashenov}},\
  }\href@noop {} {\bibfield  {journal} {\bibinfo  {journal} {Phys. Rev. Lett.}\
  }\textbf {\bibinfo {volume} {94}},\ \bibinfo {pages} {223001} (\bibinfo
  {year} {2005})}\BibitemShut {NoStop}%
\bibitem [{\citenamefont {Rischka}\ \emph {et~al.}(2020)\citenamefont
  {Rischka}, \citenamefont {Cakir}, \citenamefont {Door}, \citenamefont
  {Filianin}, \citenamefont {Harman}, \citenamefont {Huang}, \citenamefont
  {Indelicato}, \citenamefont {Keitel}, \citenamefont {K\"onig}, \citenamefont
  {Kromer}, \citenamefont {M\"uller}, \citenamefont {Novikov}, \citenamefont
  {Sch\"ussler}, \citenamefont {Schweiger}, \citenamefont {Eliseev},\ and\
  \citenamefont {Blaum}}]{Rischka2020}%
  \BibitemOpen
  \bibfield  {author} {\bibinfo {author} {\bibfnamefont {A.}~\bibnamefont
  {Rischka}}, \bibinfo {author} {\bibfnamefont {H.}~\bibnamefont {Cakir}},
  \bibinfo {author} {\bibfnamefont {M.}~\bibnamefont {Door}}, \bibinfo {author}
  {\bibfnamefont {P.}~\bibnamefont {Filianin}}, \bibinfo {author}
  {\bibfnamefont {Z.}~\bibnamefont {Harman}}, \bibinfo {author} {\bibfnamefont
  {W.~J.}\ \bibnamefont {Huang}}, \bibinfo {author} {\bibfnamefont
  {P.}~\bibnamefont {Indelicato}}, \bibinfo {author} {\bibfnamefont {C.~H.}\
  \bibnamefont {Keitel}}, \bibinfo {author} {\bibfnamefont {C.~M.}\
  \bibnamefont {K\"onig}}, \bibinfo {author} {\bibfnamefont {K.}~\bibnamefont
  {Kromer}}, \bibinfo {author} {\bibfnamefont {M.}~\bibnamefont {M\"uller}},
  \bibinfo {author} {\bibfnamefont {Y.~N.}\ \bibnamefont {Novikov}}, \bibinfo
  {author} {\bibfnamefont {R.~X.}\ \bibnamefont {Sch\"ussler}}, \bibinfo
  {author} {\bibfnamefont {C.}~\bibnamefont {Schweiger}}, \bibinfo {author}
  {\bibfnamefont {S.}~\bibnamefont {Eliseev}}, \ and\ \bibinfo {author}
  {\bibfnamefont {K.}~\bibnamefont {Blaum}},\ }\href {\doibase
  10.1103/PhysRevLett.124.113001} {\bibfield  {journal} {\bibinfo  {journal}
  {Phys. Rev. Lett.}\ }\textbf {\bibinfo {volume} {124}},\ \bibinfo {pages}
  {113001} (\bibinfo {year} {2020})}\BibitemShut {NoStop}%
\bibitem [{\citenamefont {Shabaev}\ \emph {et~al.}(2004)\citenamefont
  {Shabaev}, \citenamefont {Tupitsyn}, \citenamefont {Yerokhin}, \citenamefont
  {Plunien},\ and\ \citenamefont {Soff}}]{Shabaev2004}%
  \BibitemOpen
  \bibfield  {author} {\bibinfo {author} {\bibfnamefont {V.~M.}\ \bibnamefont
  {Shabaev}}, \bibinfo {author} {\bibfnamefont {I.~I.}\ \bibnamefont
  {Tupitsyn}}, \bibinfo {author} {\bibfnamefont {V.~A.}\ \bibnamefont
  {Yerokhin}}, \bibinfo {author} {\bibfnamefont {G.}~\bibnamefont {Plunien}}, \
  and\ \bibinfo {author} {\bibfnamefont {G.}~\bibnamefont {Soff}},\ }\href
  {\doibase 10.1103/PhysRevLett.93.130405} {\bibfield  {journal} {\bibinfo
  {journal} {Phys. Rev. Lett.}\ }\textbf {\bibinfo {volume} {93}},\ \bibinfo
  {pages} {130405} (\bibinfo {year} {2004})}\BibitemShut {NoStop}%
\bibitem [{\citenamefont {Angeli}\ and\ \citenamefont
  {Marinova}(2013)}]{Angeli2013}%
  \BibitemOpen
  \bibfield  {author} {\bibinfo {author} {\bibfnamefont {I.}~\bibnamefont
  {Angeli}}\ and\ \bibinfo {author} {\bibfnamefont {K.}~\bibnamefont
  {Marinova}},\ }\href {\doibase https://doi.org/10.1016/j.adt.2011.12.006}
  {\bibfield  {journal} {\bibinfo  {journal} {At. Data Nucl. Data Tables}\
  }\textbf {\bibinfo {volume} {99}},\ \bibinfo {pages} {69 } (\bibinfo {year}
  {2013})}\BibitemShut {NoStop}%
\bibitem [{\citenamefont {Col\`{o}}\ \emph {et~al.}(2013)\citenamefont
  {Col\`{o}}, \citenamefont {Cao}, \citenamefont {{Van Giai}},\ and\
  \citenamefont {Capelli}}]{Colo2013}%
  \BibitemOpen
  \bibfield  {author} {\bibinfo {author} {\bibfnamefont {G.}~\bibnamefont
  {Col\`{o}}}, \bibinfo {author} {\bibfnamefont {L.}~\bibnamefont {Cao}},
  \bibinfo {author} {\bibfnamefont {N.}~\bibnamefont {{Van Giai}}}, \ and\
  \bibinfo {author} {\bibfnamefont {L.}~\bibnamefont {Capelli}},\ }\href
  {\doibase https://doi.org/10.1016/j.cpc.2012.07.016} {\bibfield  {journal}
  {\bibinfo  {journal} {Comput. Phys. Commun.}\ }\textbf {\bibinfo {volume}
  {184}},\ \bibinfo {pages} {142 } (\bibinfo {year} {2013})}\BibitemShut
  {NoStop}%
\bibitem [{\citenamefont {Valuev}\ \emph {et~al.}(2020)\citenamefont {Valuev},
  \citenamefont {Harman}, \citenamefont {Keitel},\ and\ \citenamefont
  {Oreshkina}}]{Valuev2020}%
  \BibitemOpen
  \bibfield  {author} {\bibinfo {author} {\bibfnamefont {I.~A.}\ \bibnamefont
  {Valuev}}, \bibinfo {author} {\bibfnamefont {Z.}~\bibnamefont {Harman}},
  \bibinfo {author} {\bibfnamefont {C.~H.}\ \bibnamefont {Keitel}}, \ and\
  \bibinfo {author} {\bibfnamefont {N.~S.}\ \bibnamefont {Oreshkina}},\
  }\href@noop {} {} (\bibinfo {year} {2020}),\ \Eprint
  {http://arxiv.org/abs/2002.02227} {arXiv:2002.02227 [physics.atom-ph]}
  \BibitemShut {NoStop}%
\bibitem [{\citenamefont {Darwin}(1928)}]{Darwin1928}%
  \BibitemOpen
  \bibfield  {author} {\bibinfo {author} {\bibfnamefont {C.~G.}\ \bibnamefont
  {Darwin}},\ }\href@noop {} {\bibfield  {journal} {\bibinfo  {journal} {Proc.
  R. Soc. Lond. A}\ }\textbf {\bibinfo {volume} {118}},\ \bibinfo {pages} {654}
  (\bibinfo {year} {1928})}\BibitemShut {NoStop}%
\bibitem [{\citenamefont {Shabaev}(1993)}]{Shabaev1993}%
  \BibitemOpen
  \bibfield  {author} {\bibinfo {author} {\bibfnamefont {V.~M.}\ \bibnamefont
  {Shabaev}},\ }\href@noop {} {\bibfield  {journal} {\bibinfo  {journal} {J.
  Phys. B}\ }\textbf {\bibinfo {volume} {26}},\ \bibinfo {pages} {1103}
  (\bibinfo {year} {1993})}\BibitemShut {NoStop}%
\bibitem [{\citenamefont {Yerokhin}(2011)}]{Yerokhin2011}%
  \BibitemOpen
  \bibfield  {author} {\bibinfo {author} {\bibfnamefont {V.~A.}\ \bibnamefont
  {Yerokhin}},\ }\href {\doibase 10.1103/PhysRevA.83.012507} {\bibfield
  {journal} {\bibinfo  {journal} {Phys. Rev. A}\ }\textbf {\bibinfo {volume}
  {83}},\ \bibinfo {pages} {012507} (\bibinfo {year} {2011})}\BibitemShut
  {NoStop}%
\bibitem [{\citenamefont {Yerokhin}\ \emph {et~al.}(2013)\citenamefont
  {Yerokhin}, \citenamefont {Keitel},\ and\ \citenamefont
  {Harman}}]{Yerokhin2013}%
  \BibitemOpen
  \bibfield  {author} {\bibinfo {author} {\bibfnamefont {V.~A.}\ \bibnamefont
  {Yerokhin}}, \bibinfo {author} {\bibfnamefont {C.~H.}\ \bibnamefont
  {Keitel}}, \ and\ \bibinfo {author} {\bibfnamefont {Z.}~\bibnamefont
  {Harman}},\ }\href {\doibase 10.1088/0953-4075/46/24/245002} {\bibfield
  {journal} {\bibinfo  {journal} {J. Phys. B}\ }\textbf {\bibinfo {volume}
  {46}},\ \bibinfo {pages} {245002} (\bibinfo {year} {2013})}\BibitemShut
  {NoStop}%
\bibitem [{\citenamefont {Flambaum}\ \emph {et~al.}(2018)\citenamefont
  {Flambaum}, \citenamefont {Geddes},\ and\ \citenamefont
  {Viatkina}}]{Flambaum2018}%
  \BibitemOpen
  \bibfield  {author} {\bibinfo {author} {\bibfnamefont {V.~V.}\ \bibnamefont
  {Flambaum}}, \bibinfo {author} {\bibfnamefont {A.~J.}\ \bibnamefont
  {Geddes}}, \ and\ \bibinfo {author} {\bibfnamefont {A.~V.}\ \bibnamefont
  {Viatkina}},\ }\href {\doibase 10.1103/PhysRevA.97.032510} {\bibfield
  {journal} {\bibinfo  {journal} {Phys. Rev. A}\ }\textbf {\bibinfo {volume}
  {97}},\ \bibinfo {pages} {032510} (\bibinfo {year} {2018})}\BibitemShut
  {NoStop}%
\bibitem [{\citenamefont {Debierre}\ \emph {et~al.}(2019)\citenamefont
  {Debierre}, \citenamefont {Keitel},\ and\ \citenamefont
  {Harman}}]{Debierre2019}%
  \BibitemOpen
  \bibfield  {author} {\bibinfo {author} {\bibfnamefont {V.}~\bibnamefont
  {Debierre}}, \bibinfo {author} {\bibfnamefont {C.~H.}\ \bibnamefont
  {Keitel}}, \ and\ \bibinfo {author} {\bibfnamefont {Z.}~\bibnamefont
  {Harman}},\ }\href@noop {} {\bibfield  {journal} {\bibinfo  {journal}
  {submitted}\ } (\bibinfo {year} {2019})},\ \Eprint
  {http://arxiv.org/abs/arXiv:1901.06959} {arXiv:1901.06959} \BibitemShut
  {NoStop}%
\bibitem [{\citenamefont {Levinger}(1957)}]{Levinger1957}%
  \BibitemOpen
  \bibfield  {author} {\bibinfo {author} {\bibfnamefont {J.~S.}\ \bibnamefont
  {Levinger}},\ }\href {\doibase 10.1103/PhysRev.107.554} {\bibfield  {journal}
  {\bibinfo  {journal} {Phys. Rev.}\ }\textbf {\bibinfo {volume} {107}},\
  \bibinfo {pages} {554} (\bibinfo {year} {1957})}\BibitemShut {NoStop}%
\bibitem [{\citenamefont {Labzowsky}\ and\ \citenamefont
  {Nefiodov}(1994)}]{Labzowsky1994}%
  \BibitemOpen
  \bibfield  {author} {\bibinfo {author} {\bibfnamefont {L.~N.}\ \bibnamefont
  {Labzowsky}}\ and\ \bibinfo {author} {\bibfnamefont {A.~V.}\ \bibnamefont
  {Nefiodov}},\ }\href {\doibase https://doi.org/10.1016/0375-9601(94)90478-2}
  {\bibfield  {journal} {\bibinfo  {journal} {Phys. Lett. A}\ }\textbf
  {\bibinfo {volume} {188}},\ \bibinfo {pages} {371 } (\bibinfo {year}
  {1994})}\BibitemShut {NoStop}%
\bibitem [{\citenamefont {Nefiodov}\ \emph {et~al.}(1996)\citenamefont
  {Nefiodov}, \citenamefont {Labzowsky}, \citenamefont {Plunien},\ and\
  \citenamefont {Soff}}]{Nefiodov1996}%
  \BibitemOpen
  \bibfield  {author} {\bibinfo {author} {\bibfnamefont {A.~V.}\ \bibnamefont
  {Nefiodov}}, \bibinfo {author} {\bibfnamefont {L.~N.}\ \bibnamefont
  {Labzowsky}}, \bibinfo {author} {\bibfnamefont {G.}~\bibnamefont {Plunien}},
  \ and\ \bibinfo {author} {\bibfnamefont {G.}~\bibnamefont {Soff}},\ }\href
  {\doibase 10.1016/0375-9601(96)00650-0} {\bibfield  {journal} {\bibinfo
  {journal} {Phys. Lett. A}\ }\textbf {\bibinfo {volume} {222}},\ \bibinfo
  {pages} {227} (\bibinfo {year} {1996})}\BibitemShut {NoStop}%
\bibitem [{\citenamefont {Nefiodov}\ \emph {et~al.}(2003)\citenamefont
  {Nefiodov}, \citenamefont {Plunien},\ and\ \citenamefont
  {Soff}}]{Nefiodov2003}%
  \BibitemOpen
  \bibfield  {author} {\bibinfo {author} {\bibfnamefont {A.~V.}\ \bibnamefont
  {Nefiodov}}, \bibinfo {author} {\bibfnamefont {G.}~\bibnamefont {Plunien}}, \
  and\ \bibinfo {author} {\bibfnamefont {G.}~\bibnamefont {Soff}},\ }\href
  {\doibase 10.1016/S0370-2693(02)03093-9} {\bibfield  {journal} {\bibinfo
  {journal} {Phys. Lett. B}\ }\textbf {\bibinfo {volume} {552}},\ \bibinfo
  {pages} {35} (\bibinfo {year} {2003})}\BibitemShut {NoStop}%
\bibitem [{\citenamefont {Volotka}\ and\ \citenamefont
  {Plunien}(2014)}]{Volotka2014}%
  \BibitemOpen
  \bibfield  {author} {\bibinfo {author} {\bibfnamefont {A.~V.}\ \bibnamefont
  {Volotka}}\ and\ \bibinfo {author} {\bibfnamefont {G.}~\bibnamefont
  {Plunien}},\ }\href {\doibase 10.1103/PhysRevLett.113.023002} {\bibfield
  {journal} {\bibinfo  {journal} {Phys. Rev. Lett.}\ }\textbf {\bibinfo
  {volume} {113}} (\bibinfo {year} {2014}),\
  10.1103/PhysRevLett.113.023002}\BibitemShut {NoStop}%
\bibitem [{\citenamefont {Plunien}\ \emph {et~al.}(1991)\citenamefont
  {Plunien}, \citenamefont {M\"uller}, \citenamefont {Greiner},\ and\
  \citenamefont {Soff}}]{Plunien1991}%
  \BibitemOpen
  \bibfield  {author} {\bibinfo {author} {\bibfnamefont {G.}~\bibnamefont
  {Plunien}}, \bibinfo {author} {\bibfnamefont {B.}~\bibnamefont {M\"uller}},
  \bibinfo {author} {\bibfnamefont {W.}~\bibnamefont {Greiner}}, \ and\
  \bibinfo {author} {\bibfnamefont {G.}~\bibnamefont {Soff}},\ }\href {\doibase
  10.1103/PhysRevA.43.5853} {\bibfield  {journal} {\bibinfo  {journal} {Phys.
  Rev. A}\ }\textbf {\bibinfo {volume} {43}},\ \bibinfo {pages} {5853}
  (\bibinfo {year} {1991})}\BibitemShut {NoStop}%
\bibitem [{\citenamefont {{Shamsuzzoha Basunia}}(2015)}]{22}%
  \BibitemOpen
  \bibfield  {author} {\bibinfo {author} {\bibfnamefont {M.}~\bibnamefont
  {{Shamsuzzoha Basunia}}},\ }\href {\doibase
  https://doi.org/10.1016/j.nds.2015.07.002} {\bibfield  {journal} {\bibinfo
  {journal} {Nuclear Data Sheets}\ }\textbf {\bibinfo {volume} {127}},\
  \bibinfo {pages} {69 } (\bibinfo {year} {2015})}\BibitemShut {NoStop}%
\bibitem [{\citenamefont {{Shamsuzzoha Basunia}}(2013)}]{28}%
  \BibitemOpen
  \bibfield  {author} {\bibinfo {author} {\bibfnamefont {M.}~\bibnamefont
  {{Shamsuzzoha Basunia}}},\ }\href {\doibase
  https://doi.org/10.1016/j.nds.2013.10.001} {\bibfield  {journal} {\bibinfo
  {journal} {Nuclear Data Sheets}\ }\textbf {\bibinfo {volume} {114}},\
  \bibinfo {pages} {1189 } (\bibinfo {year} {2013})}\BibitemShut {NoStop}%
\bibitem [{\citenamefont {Chen}(2017)}]{40}%
  \BibitemOpen
  \bibfield  {author} {\bibinfo {author} {\bibfnamefont {J.}~\bibnamefont
  {Chen}},\ }\href {\doibase https://doi.org/10.1016/j.nds.2017.02.001}
  {\bibfield  {journal} {\bibinfo  {journal} {Nuclear Data Sheets}\ }\textbf
  {\bibinfo {volume} {140}},\ \bibinfo {pages} {1 } (\bibinfo {year}
  {2017})}\BibitemShut {NoStop}%
\bibitem [{\citenamefont {Singh}(2007)}]{64}%
  \BibitemOpen
  \bibfield  {author} {\bibinfo {author} {\bibfnamefont {B.}~\bibnamefont
  {Singh}},\ }\href {\doibase https://doi.org/10.1016/j.nds.2007.01.003}
  {\bibfield  {journal} {\bibinfo  {journal} {Nuclear Data Sheets}\ }\textbf
  {\bibinfo {volume} {108}},\ \bibinfo {pages} {197 } (\bibinfo {year}
  {2007})}\BibitemShut {NoStop}%
\bibitem [{\citenamefont {Abriola}\ \emph {et~al.}(2009)\citenamefont
  {Abriola}, \citenamefont {Bostan}, \citenamefont {Erturk}, \citenamefont
  {Fadil}, \citenamefont {Galan}, \citenamefont {Juutinen}, \citenamefont
  {Kib\'edi}, \citenamefont {Kondev}, \citenamefont {Luca}, \citenamefont
  {Negret}, \citenamefont {Nica}, \citenamefont {Pfeiffer}, \citenamefont
  {Singh}, \citenamefont {Sonzogni}, \citenamefont {Timar}, \citenamefont
  {Tuli}, \citenamefont {Venkova},\ and\ \citenamefont {Zuber}}]{84}%
  \BibitemOpen
  \bibfield  {author} {\bibinfo {author} {\bibfnamefont {D.}~\bibnamefont
  {Abriola}}, \bibinfo {author} {\bibfnamefont {M.}~\bibnamefont {Bostan}},
  \bibinfo {author} {\bibfnamefont {S.}~\bibnamefont {Erturk}}, \bibinfo
  {author} {\bibfnamefont {M.}~\bibnamefont {Fadil}}, \bibinfo {author}
  {\bibfnamefont {M.}~\bibnamefont {Galan}}, \bibinfo {author} {\bibfnamefont
  {S.}~\bibnamefont {Juutinen}}, \bibinfo {author} {\bibfnamefont
  {T.}~\bibnamefont {Kib\'edi}}, \bibinfo {author} {\bibfnamefont
  {F.}~\bibnamefont {Kondev}}, \bibinfo {author} {\bibfnamefont
  {A.}~\bibnamefont {Luca}}, \bibinfo {author} {\bibfnamefont {A.}~\bibnamefont
  {Negret}}, \bibinfo {author} {\bibfnamefont {N.}~\bibnamefont {Nica}},
  \bibinfo {author} {\bibfnamefont {B.}~\bibnamefont {Pfeiffer}}, \bibinfo
  {author} {\bibfnamefont {B.}~\bibnamefont {Singh}}, \bibinfo {author}
  {\bibfnamefont {A.}~\bibnamefont {Sonzogni}}, \bibinfo {author}
  {\bibfnamefont {J.}~\bibnamefont {Timar}}, \bibinfo {author} {\bibfnamefont
  {J.}~\bibnamefont {Tuli}}, \bibinfo {author} {\bibfnamefont {T.}~\bibnamefont
  {Venkova}}, \ and\ \bibinfo {author} {\bibfnamefont {K.}~\bibnamefont
  {Zuber}},\ }\href {\doibase https://doi.org/10.1016/j.nds.2009.10.002}
  {\bibfield  {journal} {\bibinfo  {journal} {Nuclear Data Sheets}\ }\textbf
  {\bibinfo {volume} {110}},\ \bibinfo {pages} {2815 } (\bibinfo {year}
  {2009})}\BibitemShut {NoStop}%
\bibitem [{\citenamefont {{De~Frenne}}(2009)}]{102}%
  \BibitemOpen
  \bibfield  {author} {\bibinfo {author} {\bibfnamefont {D.}~\bibnamefont
  {{De~Frenne}}},\ }\href {\doibase https://doi.org/10.1016/j.nds.2009.06.002}
  {\bibfield  {journal} {\bibinfo  {journal} {Nuclear Data Sheets}\ }\textbf
  {\bibinfo {volume} {110}},\ \bibinfo {pages} {1745 } (\bibinfo {year}
  {2009})}\BibitemShut {NoStop}%
\bibitem [{\citenamefont {Lalkovski}\ and\ \citenamefont {Kondev}(2015)}]{112}%
  \BibitemOpen
  \bibfield  {author} {\bibinfo {author} {\bibfnamefont {S.}~\bibnamefont
  {Lalkovski}}\ and\ \bibinfo {author} {\bibfnamefont {F.~G.}\ \bibnamefont
  {Kondev}},\ }\href {\doibase https://doi.org/10.1016/j.nds.2014.12.046}
  {\bibfield  {journal} {\bibinfo  {journal} {Nuclear Data Sheets}\ }\textbf
  {\bibinfo {volume} {124}},\ \bibinfo {pages} {157 } (\bibinfo {year}
  {2015})}\BibitemShut {NoStop}%
\bibitem [{\citenamefont {Johnson}\ \emph {et~al.}(2011)\citenamefont
  {Johnson}, \citenamefont {Symochko}, \citenamefont {Fadil},\ and\
  \citenamefont {Tuli}}]{142}%
  \BibitemOpen
  \bibfield  {author} {\bibinfo {author} {\bibfnamefont {T.~D.}\ \bibnamefont
  {Johnson}}, \bibinfo {author} {\bibfnamefont {D.}~\bibnamefont {Symochko}},
  \bibinfo {author} {\bibfnamefont {M.}~\bibnamefont {Fadil}}, \ and\ \bibinfo
  {author} {\bibfnamefont {J.~K.}\ \bibnamefont {Tuli}},\ }\href {\doibase
  https://doi.org/10.1016/j.nds.2011.08.002} {\bibfield  {journal} {\bibinfo
  {journal} {Nuclear Data Sheets}\ }\textbf {\bibinfo {volume} {112}},\
  \bibinfo {pages} {1949 } (\bibinfo {year} {2011})}\BibitemShut {NoStop}%
\bibitem [{\citenamefont {Nica}(2017)}]{158}%
  \BibitemOpen
  \bibfield  {author} {\bibinfo {author} {\bibfnamefont {N.}~\bibnamefont
  {Nica}},\ }\href {\doibase https://doi.org/10.1016/j.nds.2017.03.001}
  {\bibfield  {journal} {\bibinfo  {journal} {Nuclear Data Sheets}\ }\textbf
  {\bibinfo {volume} {141}},\ \bibinfo {pages} {1 } (\bibinfo {year}
  {2017})}\BibitemShut {NoStop}%
\bibitem [{\citenamefont {Reich}(2007)}]{162}%
  \BibitemOpen
  \bibfield  {author} {\bibinfo {author} {\bibfnamefont {C.~W.}\ \bibnamefont
  {Reich}},\ }\href {\doibase https://doi.org/10.1016/j.nds.2007.07.002}
  {\bibfield  {journal} {\bibinfo  {journal} {Nuclear Data Sheets}\ }\textbf
  {\bibinfo {volume} {108}},\ \bibinfo {pages} {1807 } (\bibinfo {year}
  {2007})}\BibitemShut {NoStop}%
\bibitem [{\citenamefont {Browne}\ and\ \citenamefont {Junde}(1999)}]{174}%
  \BibitemOpen
  \bibfield  {author} {\bibinfo {author} {\bibfnamefont {E.}~\bibnamefont
  {Browne}}\ and\ \bibinfo {author} {\bibfnamefont {H.}~\bibnamefont {Junde}},\
  }\href {\doibase https://doi.org/10.1006/ndsh.1999.0015} {\bibfield
  {journal} {\bibinfo  {journal} {Nuclear Data Sheets}\ }\textbf {\bibinfo
  {volume} {87}},\ \bibinfo {pages} {15 } (\bibinfo {year} {1999})}\BibitemShut
  {NoStop}%
\bibitem [{\citenamefont {Xiaolong}(2007)}]{196}%
  \BibitemOpen
  \bibfield  {author} {\bibinfo {author} {\bibfnamefont {H.}~\bibnamefont
  {Xiaolong}},\ }\href {\doibase https://doi.org/10.1016/j.nds.2007.05.001}
  {\bibfield  {journal} {\bibinfo  {journal} {Nuclear Data Sheets}\ }\textbf
  {\bibinfo {volume} {108}},\ \bibinfo {pages} {1093 } (\bibinfo {year}
  {2007})}\BibitemShut {NoStop}%
\bibitem [{\citenamefont {Martin}(2007)}]{208}%
  \BibitemOpen
  \bibfield  {author} {\bibinfo {author} {\bibfnamefont {M.~J.}\ \bibnamefont
  {Martin}},\ }\href {\doibase https://doi.org/10.1016/j.nds.2007.07.001}
  {\bibfield  {journal} {\bibinfo  {journal} {Nuclear Data Sheets}\ }\textbf
  {\bibinfo {volume} {108}},\ \bibinfo {pages} {1583 } (\bibinfo {year}
  {2007})}\BibitemShut {NoStop}%
\bibitem [{\citenamefont {Browne}\ and\ \citenamefont {Tuli}(2015)}]{238}%
  \BibitemOpen
  \bibfield  {author} {\bibinfo {author} {\bibfnamefont {E.}~\bibnamefont
  {Browne}}\ and\ \bibinfo {author} {\bibfnamefont {J.~K.}\ \bibnamefont
  {Tuli}},\ }\href {\doibase https://doi.org/10.1016/j.nds.2015.07.003}
  {\bibfield  {journal} {\bibinfo  {journal} {Nuclear Data Sheets}\ }\textbf
  {\bibinfo {volume} {127}},\ \bibinfo {pages} {191 } (\bibinfo {year}
  {2015})}\BibitemShut {NoStop}%
\bibitem [{\citenamefont {Rinker}\ and\ \citenamefont
  {Speth}(1978)}]{RINKER1978}%
  \BibitemOpen
  \bibfield  {author} {\bibinfo {author} {\bibfnamefont {G.~A.}\ \bibnamefont
  {Rinker}}\ and\ \bibinfo {author} {\bibfnamefont {J.}~\bibnamefont {Speth}},\
  }\href {\doibase https://doi.org/10.1016/0375-9474(78)90471-2} {\bibfield
  {journal} {\bibinfo  {journal} {Nucl. Phys. A}\ }\textbf {\bibinfo {volume}
  {306}},\ \bibinfo {pages} {397 } (\bibinfo {year} {1978})}\BibitemShut
  {NoStop}%
\bibitem [{\citenamefont {Yerokhin}\ and\ \citenamefont
  {Shabaev}(2015)}]{VladVladTable}%
  \BibitemOpen
  \bibfield  {author} {\bibinfo {author} {\bibfnamefont {V.~A.}\ \bibnamefont
  {Yerokhin}}\ and\ \bibinfo {author} {\bibfnamefont {V.~M.}\ \bibnamefont
  {Shabaev}},\ }\href {https://aip.scitation.org/doi/full/10.1063/1.4927487}
  {\bibfield  {journal} {\bibinfo  {journal} {J. Phys. Chem. Ref. Data}\
  }\textbf {\bibinfo {volume} {44}},\ \bibinfo {pages} {033103} (\bibinfo
  {year} {2015})}\BibitemShut {NoStop}%
\bibitem [{\citenamefont {Breit}(1928{\natexlab{b}})}]{BreitG}%
  \BibitemOpen
  \bibfield  {author} {\bibinfo {author} {\bibfnamefont {G.}~\bibnamefont
  {Breit}},\ }\href {https://www.nature.com/articles/122649a0} {\bibfield
  {journal} {\bibinfo  {journal} {Nature}\ }\textbf {\bibinfo {volume} {122}},\
  \bibinfo {pages} {649} (\bibinfo {year} {1928}{\natexlab{b}})}\BibitemShut
  {NoStop}%
\bibitem [{\citenamefont {Mohr}\ \emph {et~al.}(2016)\citenamefont {Mohr},
  \citenamefont {Newell},\ and\ \citenamefont {Taylor}}]{CODATA14}%
  \BibitemOpen
  \bibfield  {author} {\bibinfo {author} {\bibfnamefont {P.~J.}\ \bibnamefont
  {Mohr}}, \bibinfo {author} {\bibfnamefont {D.~B.}\ \bibnamefont {Newell}}, \
  and\ \bibinfo {author} {\bibfnamefont {B.~N.}\ \bibnamefont {Taylor}},\
  }\href {https://journals.aps.org/rmp/abstract/10.1103/RevModPhys.88.035009}
  {\bibfield  {journal} {\bibinfo  {journal} {Rev. Mod. Phys.}\ }\textbf
  {\bibinfo {volume} {88}},\ \bibinfo {pages} {035009} (\bibinfo {year}
  {2016})}\BibitemShut {NoStop}%
\bibitem [{\citenamefont {Yerokhin}\ and\ \citenamefont
  {Harman}(2013{\natexlab{a}})}]{VladZOneLoop}%
  \BibitemOpen
  \bibfield  {author} {\bibinfo {author} {\bibfnamefont {V.~A.}\ \bibnamefont
  {Yerokhin}}\ and\ \bibinfo {author} {\bibfnamefont {Z.}~\bibnamefont
  {Harman}},\ }\href
  {https://journals.aps.org/pra/abstract/10.1103/PhysRevA.95.060501} {\bibfield
   {journal} {\bibinfo  {journal} {Phys. Rev. A}\ }\textbf {\bibinfo {volume}
  {95}},\ \bibinfo {pages} {060501(R)} (\bibinfo {year}
  {2013}{\natexlab{a}})}\BibitemShut {NoStop}%
\bibitem [{\citenamefont {Lee}\ \emph {et~al.}(2007)\citenamefont {Lee},
  \citenamefont {Milstein}, \citenamefont {Terekhov},\ and\ \citenamefont
  {Karshenboim}}]{LeeDelbrueck}%
  \BibitemOpen
  \bibfield  {author} {\bibinfo {author} {\bibfnamefont {R.~N.}\ \bibnamefont
  {Lee}}, \bibinfo {author} {\bibfnamefont {A.~I.}\ \bibnamefont {Milstein}},
  \bibinfo {author} {\bibfnamefont {I.~S.}\ \bibnamefont {Terekhov}}, \ and\
  \bibinfo {author} {\bibfnamefont {S.~G.}\ \bibnamefont {Karshenboim}},\
  }\href {http://www.nrcresearchpress.com/doi/abs/10.1139/p07-024} {\bibfield
  {journal} {\bibinfo  {journal} {Can. J. Phys.}\ }\textbf {\bibinfo {volume}
  {85}},\ \bibinfo {pages} {541} (\bibinfo {year} {2007})}\BibitemShut
  {NoStop}%
\bibitem [{\citenamefont {Czarnecki}\ \emph {et~al.}(2000)\citenamefont
  {Czarnecki}, \citenamefont {Melnikov},\ and\ \citenamefont
  {Yelkhovsky}}]{CzarneckiEarly}%
  \BibitemOpen
  \bibfield  {author} {\bibinfo {author} {\bibfnamefont {A.}~\bibnamefont
  {Czarnecki}}, \bibinfo {author} {\bibfnamefont {K.}~\bibnamefont {Melnikov}},
  \ and\ \bibinfo {author} {\bibfnamefont {A.}~\bibnamefont {Yelkhovsky}},\
  }\href {https://journals.aps.org/pra/abstract/10.1103/PhysRevA.63.012509}
  {\bibfield  {journal} {\bibinfo  {journal} {Phys. Rev. A}\ }\textbf {\bibinfo
  {volume} {63}},\ \bibinfo {pages} {012509} (\bibinfo {year}
  {2000})}\BibitemShut {NoStop}%
\bibitem [{\citenamefont {Pachucki}\ \emph {et~al.}(2005)\citenamefont
  {Pachucki}, \citenamefont {Czarnecki}, \citenamefont {Jentschura},\ and\
  \citenamefont {Yerokhin}}]{EveryoneTwoLoop}%
  \BibitemOpen
  \bibfield  {author} {\bibinfo {author} {\bibfnamefont {K.}~\bibnamefont
  {Pachucki}}, \bibinfo {author} {\bibfnamefont {A.}~\bibnamefont {Czarnecki}},
  \bibinfo {author} {\bibfnamefont {U.~D.}\ \bibnamefont {Jentschura}}, \ and\
  \bibinfo {author} {\bibfnamefont {V.~A.}\ \bibnamefont {Yerokhin}},\ }\href
  {https://journals.aps.org/pra/abstract/10.1103/PhysRevA.72.022108} {\bibfield
   {journal} {\bibinfo  {journal} {Phys. Rev. A}\ }\textbf {\bibinfo {volume}
  {72}},\ \bibinfo {pages} {022108} (\bibinfo {year} {2005})}\BibitemShut
  {NoStop}%
\bibitem [{\citenamefont {Czarnecki}\ and\ \citenamefont
  {Szafron}(2016)}]{CzarneckiSzafron}%
  \BibitemOpen
  \bibfield  {author} {\bibinfo {author} {\bibfnamefont {A.}~\bibnamefont
  {Czarnecki}}\ and\ \bibinfo {author} {\bibfnamefont {R.}~\bibnamefont
  {Szafron}},\ }\href
  {https://journals.aps.org/pra/abstract/10.1103/PhysRevA.94.060501} {\bibfield
   {journal} {\bibinfo  {journal} {Phys. Rev. A}\ }\textbf {\bibinfo {volume}
  {94}},\ \bibinfo {pages} {060501(R)} (\bibinfo {year} {2016})}\BibitemShut
  {NoStop}%
\bibitem [{\citenamefont {Jentschura}(2009)}]{UlrichTwoLoop}%
  \BibitemOpen
  \bibfield  {author} {\bibinfo {author} {\bibfnamefont {U.~D.}\ \bibnamefont
  {Jentschura}},\ }\href
  {https://journals.aps.org/pra/abstract/10.1103/PhysRevA.79.044501} {\bibfield
   {journal} {\bibinfo  {journal} {Phys. Rev. A}\ }\textbf {\bibinfo {volume}
  {79}},\ \bibinfo {pages} {044501} (\bibinfo {year} {2009})}\BibitemShut
  {NoStop}%
\bibitem [{\citenamefont {Czarnecki}\ \emph {et~al.}(2018)\citenamefont
  {Czarnecki}, \citenamefont {Dowling}, \citenamefont {Piclum},\ and\
  \citenamefont {Szafron}}]{CzarneckiLetter}%
  \BibitemOpen
  \bibfield  {author} {\bibinfo {author} {\bibfnamefont {A.}~\bibnamefont
  {Czarnecki}}, \bibinfo {author} {\bibfnamefont {M.}~\bibnamefont {Dowling}},
  \bibinfo {author} {\bibfnamefont {J.}~\bibnamefont {Piclum}}, \ and\ \bibinfo
  {author} {\bibfnamefont {R.}~\bibnamefont {Szafron}},\ }\href
  {https://journals.aps.org/prl/abstract/10.1103/PhysRevLett.120.043203}
  {\bibfield  {journal} {\bibinfo  {journal} {Phys. Rev. Lett.}\ }\textbf
  {\bibinfo {volume} {120}},\ \bibinfo {pages} {043203} (\bibinfo {year}
  {2018})}\BibitemShut {NoStop}%
\bibitem [{\citenamefont {Yerokhin}\ and\ \citenamefont
  {Harman}(2013{\natexlab{b}})}]{VladZVP}%
  \BibitemOpen
  \bibfield  {author} {\bibinfo {author} {\bibfnamefont {V.~A.}\ \bibnamefont
  {Yerokhin}}\ and\ \bibinfo {author} {\bibfnamefont {Z.}~\bibnamefont
  {Harman}},\ }\href
  {https://journals.aps.org/pra/abstract/10.1103/PhysRevA.88.042502} {\bibfield
   {journal} {\bibinfo  {journal} {Phys. Rev. A}\ }\textbf {\bibinfo {volume}
  {88}},\ \bibinfo {pages} {042502} (\bibinfo {year}
  {2013}{\natexlab{b}})}\BibitemShut {NoStop}%
\bibitem [{\citenamefont {Sikora}(2018)}]{BastianPhD}%
  \BibitemOpen
  \bibfield  {author} {\bibinfo {author} {\bibfnamefont {B.}~\bibnamefont
  {Sikora}},\ }\emph {\bibinfo {title} {{Quantum field theory of the $g$-factor
  of bound systems}}},\ \href@noop {} {Ph.D. thesis},\ \bibinfo  {school}
  {Heidelberg University} (\bibinfo {year} {2018})\BibitemShut {NoStop}%
\bibitem [{\citenamefont {Melnikov}\ and\ \citenamefont {van
  Ritbergen}(2000)}]{ThreeLoopSlope}%
  \BibitemOpen
  \bibfield  {author} {\bibinfo {author} {\bibfnamefont {K.}~\bibnamefont
  {Melnikov}}\ and\ \bibinfo {author} {\bibfnamefont {T.}~\bibnamefont {van
  Ritbergen}},\ }\href
  {https://journals.aps.org/prl/abstract/10.1103/PhysRevLett.84.1673}
  {\bibfield  {journal} {\bibinfo  {journal} {Phys. Rev. Lett.}\ }\textbf
  {\bibinfo {volume} {84}},\ \bibinfo {pages} {1673} (\bibinfo {year}
  {2000})}\BibitemShut {NoStop}%
\bibitem [{\citenamefont {Karshenboim}\ and\ \citenamefont
  {Shelyuto}(2019)}]{ThreeLoopRad}%
  \BibitemOpen
  \bibfield  {author} {\bibinfo {author} {\bibfnamefont {S.~G.}\ \bibnamefont
  {Karshenboim}}\ and\ \bibinfo {author} {\bibfnamefont {V.~A.}\ \bibnamefont
  {Shelyuto}},\ }\href
  {https://journals.aps.org/pra/abstract/10.1103/PhysRevA.100.032513}
  {\bibfield  {journal} {\bibinfo  {journal} {Phys. Rev. A}\ }\textbf {\bibinfo
  {volume} {100}},\ \bibinfo {pages} {032513} (\bibinfo {year}
  {2019})}\BibitemShut {NoStop}%
\bibitem [{\citenamefont {Shabaev}\ and\ \citenamefont
  {Yerokhin}(2002)}]{RecoilAllOrders}%
  \BibitemOpen
  \bibfield  {author} {\bibinfo {author} {\bibfnamefont {V.~M.}\ \bibnamefont
  {Shabaev}}\ and\ \bibinfo {author} {\bibfnamefont {V.~A.}\ \bibnamefont
  {Yerokhin}},\ }\href
  {https://journals.aps.org/prl/abstract/10.1103/PhysRevLett.88.091801}
  {\bibfield  {journal} {\bibinfo  {journal} {Phys. Rev. Lett.}\ }\textbf
  {\bibinfo {volume} {88}},\ \bibinfo {pages} {091801} (\bibinfo {year}
  {2002})}\BibitemShut {NoStop}%
\bibitem [{\citenamefont {Pachucki}(2008)}]{HOMass}%
  \BibitemOpen
  \bibfield  {author} {\bibinfo {author} {\bibfnamefont {K.}~\bibnamefont
  {Pachucki}},\ }\href
  {https://journals.aps.org/pra/abstract/10.1103/PhysRevA.78.012504} {\bibfield
   {journal} {\bibinfo  {journal} {Phys. Rev. A}\ }\textbf {\bibinfo {volume}
  {78}},\ \bibinfo {pages} {012504} (\bibinfo {year} {2008})}\BibitemShut
  {NoStop}%
\bibitem [{\citenamefont {Shabaev}\ \emph {et~al.}(2015)\citenamefont
  {Shabaev}, \citenamefont {Glazov}, \citenamefont {Plunien},\ and\
  \citenamefont {Volotka}}]{ShabaevReview}%
  \BibitemOpen
  \bibfield  {author} {\bibinfo {author} {\bibfnamefont {V.~M.}\ \bibnamefont
  {Shabaev}}, \bibinfo {author} {\bibfnamefont {D.~A.}\ \bibnamefont {Glazov}},
  \bibinfo {author} {\bibfnamefont {G.}~\bibnamefont {Plunien}}, \ and\
  \bibinfo {author} {\bibfnamefont {A.~V.}\ \bibnamefont {Volotka}},\ }\href
  {https://doi.org/10.1063/1.4921299} {\bibfield  {journal} {\bibinfo
  {journal} {J. Phys. Chem. Ref. Data}\ }\textbf {\bibinfo {volume} {44}},\
  \bibinfo {pages} {031205} (\bibinfo {year} {2015})},\ \Eprint
  {http://arxiv.org/abs/https://doi.org/10.1063/1.4921299}
  {https://doi.org/10.1063/1.4921299} \BibitemShut {NoStop}%
\bibitem [{\citenamefont {Sikora}\ \emph {et~al.}(2020)\citenamefont {Sikora},
  \citenamefont {Yerokhin}, \citenamefont {Oreshkina}, \citenamefont {Cakir},
  \citenamefont {Keitel},\ and\ \citenamefont {Harman}}]{Sikora2020}%
  \BibitemOpen
  \bibfield  {author} {\bibinfo {author} {\bibfnamefont {B.}~\bibnamefont
  {Sikora}}, \bibinfo {author} {\bibfnamefont {V.~A.}\ \bibnamefont
  {Yerokhin}}, \bibinfo {author} {\bibfnamefont {N.~S.}\ \bibnamefont
  {Oreshkina}}, \bibinfo {author} {\bibfnamefont {H.}~\bibnamefont {Cakir}},
  \bibinfo {author} {\bibfnamefont {C.~H.}\ \bibnamefont {Keitel}}, \ and\
  \bibinfo {author} {\bibfnamefont {Z.}~\bibnamefont {Harman}},\ }\href
  {\doibase 10.1103/PhysRevResearch.2.012002} {\bibfield  {journal} {\bibinfo
  {journal} {Phys. Rev. Research}\ }\textbf {\bibinfo {volume} {2}},\ \bibinfo
  {pages} {012002} (\bibinfo {year} {2020})}\BibitemShut {NoStop}%
\bibitem [{\citenamefont {Oreshkina}\ \emph {et~al.}(2020)\citenamefont
  {Oreshkina}, \citenamefont {Cakir}, \citenamefont {Sikora}, \citenamefont
  {Yerokhin}, \citenamefont {Debierre}, \citenamefont {Harman},\ and\
  \citenamefont {Keitel}}]{Oreshkina2020}%
  \BibitemOpen
  \bibfield  {author} {\bibinfo {author} {\bibfnamefont {N.~S.}\ \bibnamefont
  {Oreshkina}}, \bibinfo {author} {\bibfnamefont {H.}~\bibnamefont {Cakir}},
  \bibinfo {author} {\bibfnamefont {B.}~\bibnamefont {Sikora}}, \bibinfo
  {author} {\bibfnamefont {V.~A.}\ \bibnamefont {Yerokhin}}, \bibinfo {author}
  {\bibfnamefont {V.}~\bibnamefont {Debierre}}, \bibinfo {author}
  {\bibfnamefont {Z.}~\bibnamefont {Harman}}, \ and\ \bibinfo {author}
  {\bibfnamefont {C.~H.}\ \bibnamefont {Keitel}},\ }\href {\doibase
  10.1103/PhysRevA.101.032511} {\bibfield  {journal} {\bibinfo  {journal}
  {Phys. Rev. A}\ }\textbf {\bibinfo {volume} {101}},\ \bibinfo {pages}
  {032511} (\bibinfo {year} {2020})}\BibitemShut {NoStop}%
\bibitem [{\citenamefont {Sturm}\ \emph {et~al.}(2019)\citenamefont {Sturm},
  \citenamefont {Arapoglou}, \citenamefont {Egl}, \citenamefont {H{\"o}cker},
  \citenamefont {Kraemer}, \citenamefont {Sailer}, \citenamefont {Tu},
  \citenamefont {Weigel}, \citenamefont {Wolf}, \citenamefont {{Crespo
  L{\'o}pez-Urrutia}},\ and\ \citenamefont {Blaum}}]{Sturm2019}%
  \BibitemOpen
  \bibfield  {author} {\bibinfo {author} {\bibfnamefont {S.}~\bibnamefont
  {Sturm}}, \bibinfo {author} {\bibfnamefont {I.}~\bibnamefont {Arapoglou}},
  \bibinfo {author} {\bibfnamefont {A.}~\bibnamefont {Egl}}, \bibinfo {author}
  {\bibfnamefont {M.}~\bibnamefont {H{\"o}cker}}, \bibinfo {author}
  {\bibfnamefont {S.}~\bibnamefont {Kraemer}}, \bibinfo {author} {\bibfnamefont
  {T.}~\bibnamefont {Sailer}}, \bibinfo {author} {\bibfnamefont
  {B.}~\bibnamefont {Tu}}, \bibinfo {author} {\bibfnamefont {A.}~\bibnamefont
  {Weigel}}, \bibinfo {author} {\bibfnamefont {R.}~\bibnamefont {Wolf}},
  \bibinfo {author} {\bibfnamefont {J.}~\bibnamefont {{Crespo
  L{\'o}pez-Urrutia}}}, \ and\ \bibinfo {author} {\bibfnamefont
  {K.}~\bibnamefont {Blaum}},\ }\href@noop {} {\bibfield  {journal} {\bibinfo
  {journal} {Eur. Phys. J. Spec. Top.}\ }\textbf {\bibinfo {volume} {227}},\
  \bibinfo {pages} {1425} (\bibinfo {year} {2019})}\BibitemShut {NoStop}%
\bibitem [{\citenamefont {Nauta}\ \emph {et~al.}(2017)\citenamefont {Nauta},
  \citenamefont {Borodin}, \citenamefont {Ledwa}, \citenamefont {Stark},
  \citenamefont {Schwarz}, \citenamefont {Schm{\"o}ger}, \citenamefont {Micke},
  \citenamefont {{Crespo López-Urrutia}},\ and\ \citenamefont
  {Pfeifer}}]{Nauta2017}%
  \BibitemOpen
  \bibfield  {author} {\bibinfo {author} {\bibfnamefont {J.}~\bibnamefont
  {Nauta}}, \bibinfo {author} {\bibfnamefont {A.}~\bibnamefont {Borodin}},
  \bibinfo {author} {\bibfnamefont {H.~B.}\ \bibnamefont {Ledwa}}, \bibinfo
  {author} {\bibfnamefont {J.}~\bibnamefont {Stark}}, \bibinfo {author}
  {\bibfnamefont {M.}~\bibnamefont {Schwarz}}, \bibinfo {author} {\bibfnamefont
  {L.}~\bibnamefont {Schm{\"o}ger}}, \bibinfo {author} {\bibfnamefont
  {P.}~\bibnamefont {Micke}}, \bibinfo {author} {\bibfnamefont {J.~R.}\
  \bibnamefont {{Crespo López-Urrutia}}}, \ and\ \bibinfo {author}
  {\bibfnamefont {T.}~\bibnamefont {Pfeifer}},\ }\href {\doibase
  https://doi.org/10.1016/j.nimb.2017.04.077} {\bibfield  {journal} {\bibinfo
  {journal} {Nuclear Instruments and Methods in Physics Research Section B:
  Beam Interactions with Materials and Atoms}\ }\textbf {\bibinfo {volume}
  {408}},\ \bibinfo {pages} {285 } (\bibinfo {year} {2017})},\ \bibinfo {note}
  {{Proceedings of the 18th International Conference on the Physics of Highly
  Charged Ions (HCI-2016), Kielce, Poland, 11-16 September 2016}}\BibitemShut
  {NoStop}%
\bibitem [{\citenamefont {Schm{\"o}ger}\ \emph {et~al.}(2015)\citenamefont
  {Schm{\"o}ger}, \citenamefont {Versolato}, \citenamefont {Schwarz},
  \citenamefont {Kohnen}, \citenamefont {Windberger}, \citenamefont {Piest},
  \citenamefont {Feuchtenbeiner}, \citenamefont {Pedregosa-Gutierrez},
  \citenamefont {Leopold}, \citenamefont {Micke}, \citenamefont {Hansen},
  \citenamefont {Baumann}, \citenamefont {Drewsen}, \citenamefont {Ullrich},
  \citenamefont {Schmidt},\ and\ \citenamefont {{Crespo
  L{\'o}pez-Urrutia}}}]{Schmoeger2015}%
  \BibitemOpen
  \bibfield  {author} {\bibinfo {author} {\bibfnamefont {L.}~\bibnamefont
  {Schm{\"o}ger}}, \bibinfo {author} {\bibfnamefont {O.~O.}\ \bibnamefont
  {Versolato}}, \bibinfo {author} {\bibfnamefont {M.}~\bibnamefont {Schwarz}},
  \bibinfo {author} {\bibfnamefont {M.}~\bibnamefont {Kohnen}}, \bibinfo
  {author} {\bibfnamefont {A.}~\bibnamefont {Windberger}}, \bibinfo {author}
  {\bibfnamefont {B.}~\bibnamefont {Piest}}, \bibinfo {author} {\bibfnamefont
  {S.}~\bibnamefont {Feuchtenbeiner}}, \bibinfo {author} {\bibfnamefont
  {J.}~\bibnamefont {Pedregosa-Gutierrez}}, \bibinfo {author} {\bibfnamefont
  {T.}~\bibnamefont {Leopold}}, \bibinfo {author} {\bibfnamefont
  {P.}~\bibnamefont {Micke}}, \bibinfo {author} {\bibfnamefont {A.~K.}\
  \bibnamefont {Hansen}}, \bibinfo {author} {\bibfnamefont {T.~M.}\
  \bibnamefont {Baumann}}, \bibinfo {author} {\bibfnamefont {M.}~\bibnamefont
  {Drewsen}}, \bibinfo {author} {\bibfnamefont {J.}~\bibnamefont {Ullrich}},
  \bibinfo {author} {\bibfnamefont {P.~O.}\ \bibnamefont {Schmidt}}, \ and\
  \bibinfo {author} {\bibfnamefont {J.~R.}\ \bibnamefont {{Crespo
  L{\'o}pez-Urrutia}}},\ }\href {\doibase 10.1126/science.aaa2960} {\bibfield
  {journal} {\bibinfo  {journal} {Science}\ }\textbf {\bibinfo {volume}
  {347}},\ \bibinfo {pages} {1233} (\bibinfo {year} {2015})}\BibitemShut
  {NoStop}%
\bibitem [{\citenamefont {Micke}\ \emph {et~al.}(2020)\citenamefont {Micke},
  \citenamefont {Leopold}, \citenamefont {King}, \citenamefont {Benkler},
  \citenamefont {Spie{\ss}}, \citenamefont {Schm{\"o}ger}, \citenamefont
  {Schwarz}, \citenamefont {{Crespo L{\'o}pez-Urrutia}},\ and\ \citenamefont
  {Schmidt}}]{Micke2020}%
  \BibitemOpen
  \bibfield  {author} {\bibinfo {author} {\bibfnamefont {P.}~\bibnamefont
  {Micke}}, \bibinfo {author} {\bibfnamefont {T.}~\bibnamefont {Leopold}},
  \bibinfo {author} {\bibfnamefont {S.~A.}\ \bibnamefont {King}}, \bibinfo
  {author} {\bibfnamefont {E.}~\bibnamefont {Benkler}}, \bibinfo {author}
  {\bibfnamefont {L.~J.}\ \bibnamefont {Spie{\ss}}}, \bibinfo {author}
  {\bibfnamefont {L.}~\bibnamefont {Schm{\"o}ger}}, \bibinfo {author}
  {\bibfnamefont {M.}~\bibnamefont {Schwarz}}, \bibinfo {author} {\bibfnamefont
  {J.~R.}\ \bibnamefont {{Crespo L{\'o}pez-Urrutia}}}, \ and\ \bibinfo {author}
  {\bibfnamefont {P.~O.}\ \bibnamefont {Schmidt}},\ }\href@noop {} {\bibfield
  {journal} {\bibinfo  {journal} {Nature}\ }\textbf {\bibinfo {volume} {578}},\
  \bibinfo {pages} {60} (\bibinfo {year} {2020})}\BibitemShut {NoStop}%
\end{thebibliography}
\end{document}